\documentclass[10pt,journal,compsoc]{IEEEtran}
\ifCLASSOPTIONcompsoc
  \usepackage[nocompress]{cite}
\else
  \usepackage{cite}
\fi
\ifCLASSINFOpdf
\else
\fi
\usepackage{url}
\usepackage{color}
\usepackage[sort&compress, numbers]{natbib}
\usepackage{graphicx}
\usepackage{adjustbox}
\usepackage{amsmath}
\usepackage{algorithmic}
\usepackage{caption}
\usepackage{multirow}
\usepackage{subcaption}
\usepackage{algorithm}
\usepackage{adjustbox}
\usepackage{amsmath,bm}

\def\degree{${}^{\circ}$}

\usepackage{amsmath,amsfonts,bm}

\def\eqref#1{equation~\ref{#1}}
\def\1{\bm{1}}

\DeclareMathAlphabet{\mathsfit}{\encodingdefault}{\sfdefault}{m}{sl}
\SetMathAlphabet{\mathsfit}{bold}{\encodingdefault}{\sfdefault}{bx}{n}

\begin{document}

\title{Adaptive Cross-Layer Attention for Image Restoration}

\author{Yancheng~Wang, Student Member, IEEE, \\
        Ning~Xu, Senior Member, IEEE,
        and~Yingzhen~Yang, Member, IEEE% <-this % stops a space
\IEEEcompsocitemizethanks{\IEEEcompsocthanksitem Yancheng~Wang and Yingzhen Yang are with School of Computing and
Augmented Intelligence, Arizona State University, Tempe, AZ, 85281.\protect\\
E-mail: ywan1053@asu.edu,yingzhen.yang@asu.edu
\IEEEcompsocthanksitem Ning Xu is with Kuaishou Technology. \protect\\
E-mail: ningxu01@gmail.com
}% <-this % stops an unwanted space
}

\IEEEtitleabstractindextext{%
\begin{abstract}
Non-local attention module has been proven to be crucial for image restoration. Conventional non-local attention processes features of each layer separately, so it risks missing correlation between features among different layers. To address this problem, we aim to design attention modules that aggregate information from different layers. Instead of finding correlated key pixels within the same layer, each query pixel is encouraged to attend to key pixels at multiple previous layers of the network. In order to efficiently embed such attention design into neural network backbones, we propose a novel Adaptive Cross-Layer Attention (ACLA) module. Two adaptive designs are proposed for ACLA: (1) adaptively selecting the keys for non-local attention at each layer; (2) automatically searching for the insertion locations for ACLA modules. By these two adaptive designs, ACLA dynamically selects a flexible number of keys to be aggregated for non-local attention at previous layer while maintaining a compact neural network with compelling performance. Extensive experiments on image restoration tasks, including single image super-resolution, image denoising, image demosaicing, and image compression artifacts reduction, validate the effectiveness and efficiency of ACLA.
The code of ACLA is available at \url{https://github.com/SDL-ASU/ACLA}.
\end{abstract}
}
\maketitle

\IEEEdisplaynontitleabstractindextext

% \fi
\IEEEpeerreviewmaketitle

\IEEEraisesectionheading{\section{Introduction}}
\IEEEPARstart{I}{mage} restoration algorithms aim to recover a high-quality image from a contaminated input image by solving an ill-posed image restoration problem. There are various image restoration tasks depending on the type of corruption, such as image denoising \cite{zhang2019residual,liu2018non}, demosaicing \cite{zhang2017learning,zhang2019residual}, single image super-resolution \cite{fan2019scale,lai2017deep,tai2017memnet}, and image compression artifacts reduction\cite{zhang2017beyond}. To restore corrupted information from the contaminated image, a variety of image priors \cite{buades2005non,zoran2011learning,zontak2013separating} were proposed.

Recently, image restoration methods based on deep neural networks have achieved great success. Inspired by the widely used non-local prior, most recent approaches based on neural networks \cite{zhang2019residual,liu2018non} adapt non-local attention into their neural network to enhance the representation learning, following the non-local neural networks \cite{wang2018non}. In a non-local block, a response is calculated as a weighted sum over all pixel-wise features on the feature map to account for long-range information. Such a module was initially designed for high-level recognition tasks such as image classification, and it has been proven to be beneficial for low-level vision tasks~\cite{zhang2019residual,liu2018non}.

Though attention modules have been shown to be effective in boosting performance, most attention modules only explore the correlation among features at the same layer. Actually, features at different intermediate layers encode variant information at different scales and might be helpful to augment the information used in recovering the high-quality image. Motivated by the potential benefit of exploring feature correlation across intermediate layers, Holistic Attention Network (HAN)~\cite{niu2020single} is proposed to find the interrelationship among features at hierarchical levels with a Layer Attention Module (LAM). However, LAM assigns a single importance weight to all features at the same layer and neglects the difference in spatial positions of these features. Recent research in omnidirectional representation~\cite{tay2021omninet} suggests that exploring the relationship among features at different layers can benefit the representation learning of neural networks. Nevertheless, calculating correlation among features at hierarchical layers is computationally expensive due to the quadratic complexity of dot product attention. The complexity of such cross-layer attention design is increased from $(HW)^2L$ to $(HWL)^2$, where $H,W$ are the height and width of the feature map and $L$ is the number of layers. To handle the limitations of the current attention modules, we propose a novel Adaptive Cross-Layer Attention (ACLA) module for various image restoration tasks.

\vspace{-2mm}
\subsection{Contributions}

Our contributions are presented as follows.

First, in order to address the limitation caused by only referring to keys within the same layer in most existing attention modules, ACLA module searches for keys across different layers for each query feature, and each query only attends to a small set of keys at different layers. We name the layers where keys are attended to by a query feature the referred layers of that query.

Second, ACLA selects an adaptive number of keys at each layer for each query, and searches for the optimal insert positions. The two adaptive designs, the adaptive key selection and search for insert positions, are designed for both efficiency and effectiveness of the attention mechanism and they are inspired by neural architecture search. Because a query feature only attend to keys at previous layers where ACLA modules are available, ACLA enables automatic search for referred layers for each query.

To demonstrate the effectiveness of the two adaptive designs, we deploy ACLA modules on a commonly used neural network model, EDSR~\cite{EDSR}, for image restoration. Extensive experiments on single image super-resolution, image denoising, image compression artifacts reduction, and image demosaicing demonstrate the effectiveness of our approach. Moreover,
comprehensive ablation studies are conducted in Section~\ref{sec:ablation_study} to explain the superior performance of ACLA over its competing methods, as well as the superiority of the two adaptive designs for ACLA. In particular, ACLA is compared to the  competing attention modules in Section~\ref{sec:ablation-sota-attention}, and ablation study for the two adapative designes of ACLA is performed in Section~\ref{sec:ablation-two-adaptive-designs}. The benefit of automatic search for referred layers is discussed in Section~\ref{sec:ablation-referred-layers}. The visualization of keys selected by ACLA are illustrated in Figure~\ref{fig:key_sample_vis} and Figure~\ref{fig:more_visualization}.

This paper is organized as follows. Section~\ref{sec:related-works} introduces the related works including neural networks for image restoration, attention mechanism, and neural architecture search. The detailed formulation of ACLA is introduced in Section~\ref{sec:ACLA}. The experimental results of ACLA for various image restoration tasks and the ablation studies of ACLA are reported in Section~\ref{sec:experiments}. We conclude the paper in Section~\ref{sec:conclusion}.

\section{Related Works}
\label{sec:related-works}
\subsection{Neural Networks for Image Restoration}
Adopting neural networks for image restoration has achieved great success by utilizing their power in representation. ARCNN \cite{dong2015compression} was first proposed to use CNN for compression artifacts reduction. Later, DnCNN \cite{zhang2017beyond} uses residual learning and batch normalization to boost the performance of CNN for image denoising. In IRCNN \cite{zhang2017learning}, a learned set of CNNs are used as denoising prior for other image restoration tasks. For single image super-resolution\cite{lai2017deep, zhang2018residual, haris2018deep, fan2019scale}, even more efforts have been devoted to designing advanced architectures and learning methods. For example, RDN \cite{zhang2018residual} and CARN \cite{ahn2018fast} fuse low-level and high-level features with dense connections to provide richer information and details for reconstructing. Recently, non-local attention \cite{liu2018non,dai2019second, mei2020pyramid} is also used to further boost the performance of CNN for image restoration.

\subsection{Attention Mechanism}
Attention mechanism has been applied to many computer vision tasks, such as image captioning \cite{xu2015show,chen2017sca} and image classification \cite{hu2018squeeze, wang2017residual}. Non-local attention \cite{wang2018non} was first proposed to capture long-range dependencies for high-level recognition tasks. Recently, several works have proposed to leverage non-local attention for low-level vision tasks. In NLRN \cite{liu2018non} a recurrent neural network is proposed to incorporate non-local attention. RNAN \cite{zhang2019residual} proposed a residual local and non-local mask branch to obtain non-local mixed attention. RCAN \cite{zhang2018image} exploits the interdependencies among feature channels by generating different attention for each channel-wise feature. HAN~\cite{niu2020single} is proposed to find interrelationships among features at hierarchical levels with a layer attention module. Besides, some recent works attempt to explore the benefits of transformer-based models for image restoration. IPT \cite{IPT} is proposed to solve various restoration problems in a multi-task learning framework based on visual Transformer. SwinIR \cite{liang2021swinir} adopts the architecture of Swin Transformer. However, compared with methods using CNN architecture, transformer-based image restoration methods usually use large datasets for training. Specifically, IPT uses ImageNet to pretrain the model. SwinIR adapts a combination of four datasets consisting of over 8000 high-quality images as the training set for the tasks of denoising and compression artifact reduction.

\subsection{Neural Architecture Search}
Neural Architecture Search (NAS) has attracted lots of attention recently. Early works of NAS adopt heuristic methods such as reinforcement learning~\cite{zoph2016neural} and evolutionary algorithm~\cite{xie2017genetic}. The search process with such methods requires huge computational resources. Recently, various strategies are designed to reduce the expensive costs including weight sharing~\cite{pham2018efficient}, progressive search~\cite{liu2018progressive} and one-shot search~\cite{liu2018darts,xie2018snas}. For example, DARTS~\cite{liu2018darts} firstly relaxes the search space to be continuous and conducts the differentiable search. The architecture parameters and network weights are trained simultaneously by gradient descent to reduce the search time.

Despite the success of NAS methods for classification, dense prediction tasks such as semantic image segmentation and image restoration, usually demand more complicated network architectures. Some recent works have been devoted to exploring hierarchical search space for dense prediction tasks. For example, Auto-DeepLab \cite{liu2019auto} introduces a hierarchical search space for semantic image segmentation. DCNAS \cite{zhang2021dcnas} build a densely connected search space to extract multi-level information. HNAS \cite{guo2020hierarchical} also adopts a hierarchical search space for single image super-resolution.
\begin{figure*}[!htbp]
\begin{center}
  \includegraphics[width=0.9\linewidth]{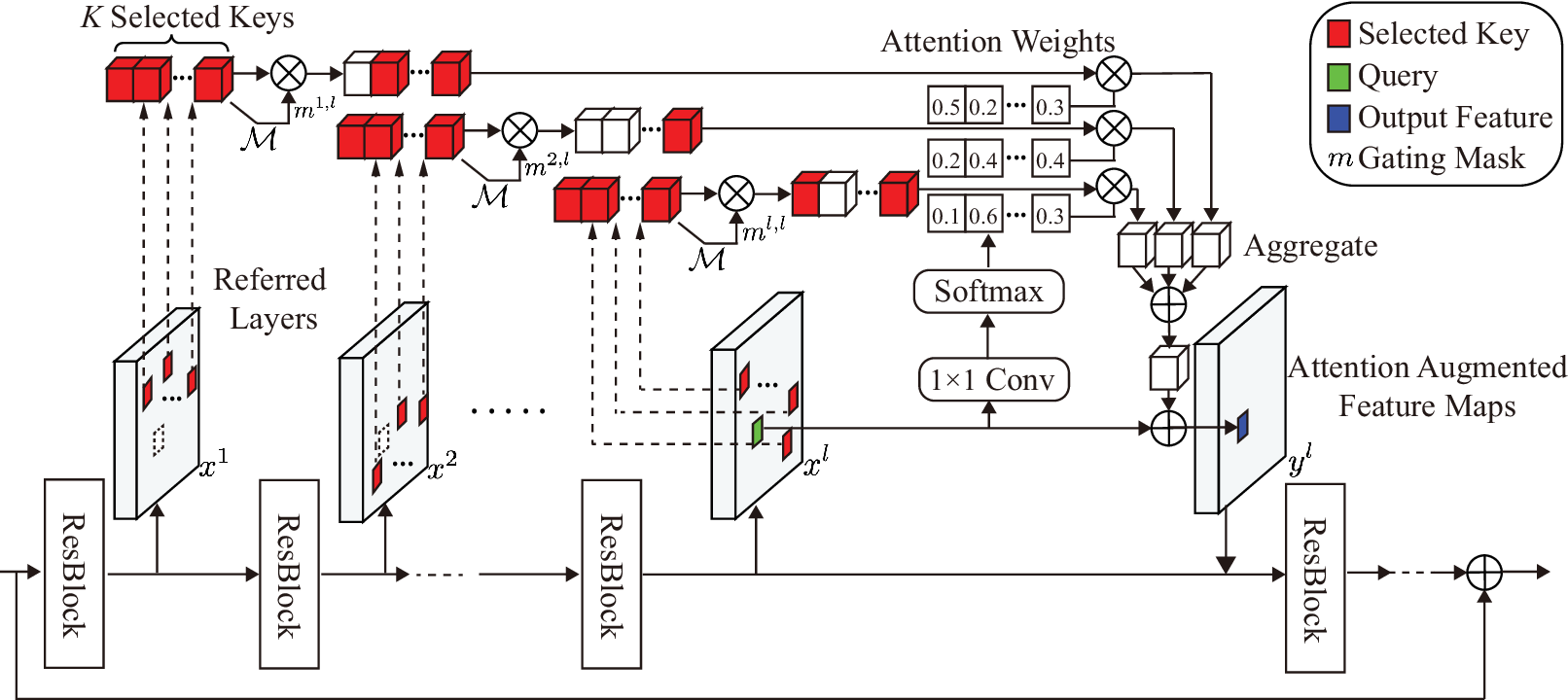}
\end{center}
\caption{Illustration of Adaptive Key Selection in an Adaptive Cross-Layer Attention (ACLA) module. For each query pixel, ACLA first selects a fixed number, $K$, of key features from each referred layer $x^j$, with $j$ from $\{1,...,l\}$. The locations for the selected keys are obtained by applying a $1\times1$ convolution layer on the query feature. Next, we apply the masking unit $\mathcal{M}$ from Equation(\ref{eq:mask_unit}) to the selected keys to generate the gating masks $\{m^{j,l}\}_{j=1}^{l}$. By multiplying the gating masks on the selected keys, we achieve adaptive key selection from each referred layer. A convolution layer and Softmax are applied to the query feature to generate attention weights for the selected keys. Weighted by the attention weights, the features of the selected keys are aggregated to the query feature to generate the output of the ACLA module.}
\label{fig:ACLA}
\end{figure*}

\section{ACLA: Adaptive Cross-Layer Attention}
\label{sec:ACLA}

We detail the formulation of ACLA in the section. The vanilla non-local attention and the proposed adaptive cross-layer attention are introduced in Section~\ref{sec:CLA}, and the search for insert positions of ACLA modules is described in Section~\ref{sec:ACLA-insert-position}.

\subsection{Cross-Layer Attention}
\label{sec:CLA}
\textbf{Vanilla Non-Local Attention.~}
Non-Local (NL) attention \cite{wang2018non} is designed to integrate the self-attention mechanism into convolutional neural networks for computer vision tasks. It is usually applied on an input feature map $x\in \mathbb{R}^{H\times W \times C}$ to explore self-similarities among all spatial positions. We reshape $x$ to $N \times C$, $N=H\times W$, where $H$, $W$, and $C$ are the height, width, and channel number of the input feature map $X$. A generic NL attention can be formulated as
\begin{equation} \label{eq:original_nl}
    y_i = \frac{1}{\mathcal{C}(x)} \sum_{n=1}^{N}f(x_i, x_n)g(x_n),
\end{equation}
where $i$ indexes the spatial position of feature maps. $y$ is the output of NL attention with the same size as $x$. $f(x_i, x_n)$ is the pairwise affinity between the query feature $x_i$ and its key feature $x_n$. $g(x_n)$ computes an embedding of feature $x_n$. $C(x)$ is a normalization term.

NL attention is usually wrapped into a non-local block \cite{wang2018non} with a residual connection from the input feature $x$. The mathematical formulation is given as
\begin{equation} \label{eq:residual}
    z = h(y) + x,
\end{equation}
where $h$ denotes a learnable feature transformation, which takes the output of non-local attention (\ref{eq:original_nl}) as input.
% \textbf{Network Structure for Image Restoration}

\textbf{Adaptive Cross-Layer Attention.~}
To search for keys from different layers for each query feature, we first adapt NL attention in Equation~(\ref{eq:original_nl}) to a cross-layer design, such that features from different layers are regarded as keys.
% To search for keys from different layers, we first concatenate features from different layers into a spatially $3$D tensor.

In the sequel, the superscript indicates the index of a layer, and the subscript indicates spatial location. Suppose that $x^{i}$ is the output of the $i$-th layer in a CNN backbone for image restoration, where $i\in \{1,\cdots, L\}$ and $L$ is the number of layers. A vanilla Cross-Layer Non-Local (CLNL) attention is formulated as
\begin{equation} \label{eq:cl_nl}
    y_i^{j} = \frac{1}{\mathcal{C}(x^j)} \sum_{l=1}^{j} \sum_{n=1}^{N}f(x_i^j, x_n^l)g(x_n^l),
\end{equation}
% where the superscripts $j,l$ index the layer and the subscripts $i,n$ index the spatial locations of features. $y,x$ denote the output feature and input feature respectively.
where the subscripts $i,n$ index the spatial locations of features, the superscripts $j,l$ are the layer indices, and $y,x$ denote the output feature and input feature respectively.
With such adaption, relationships among features across different layers can be captured. However, given the quadratic complexity of correlation computation, the complexity of CLNL is increased from $N^2L$ to $(NL)^2$. In order to mitigate the expensive inference cost, we propose to select only a small number, $K$, of key features from each referred layer for the attention module, where $K\ll N$.
% To select $K$ such key features from each referred layer,
We find the locations of selected keys from each referred layer by learning their offsets from the position of the query feature with the deformable convolution proposed in DCN~\cite{dai2017deformable}.
% By applying the deformable mechanism proposed in deformable convolution (DCN)~\cite{dai2017deformable},  .
As a result, the key features $\{x_n^l\}_{n=1}^{N}$ in the vanilla CLNL are replaced by $\{x^l(p_i+\Delta p_{ik})\}_{k=1}^{K}$,
% \begin{equation} \label{eq:ori_nl}
%     y_i^{j} = \frac{1}{\mathcal{C}(x^j)} \sum_{l=1}^{j} \sum_{k=1}^{K}f(x_i^j, x^l(p_i+\Delta p_{ik}))g(x^l(p_i+\Delta p_{ik})),
% \end{equation}
where $k$ indexes the sampled keys, and $l$ indexes the referred layer. $p_i$ denotes the 2D spatial position of the query feature $x_i^j$ in the feature map, and $\Delta p_{ik}$ is the 2-d offset from the position $p_i$ to the position of corresponding sampled key. As $p_i + \Delta p_{ik}$ can be fractional, bilinear interpolation is used as in \cite{dai2017deformable} to compute $x(p_i + \Delta p_{ik})$. To further reduce the computational complexity, we generate the attention weights from the query feature alone by $f(x_i^j)$, where $f$ is a $1\times 1$ convolution followed by a Softmax operation in our work.

% Thus, our proposed CLA can be simplified as
% \begin{equation} \label{eq:final_csnl}
%     y_i^{j} = \frac{1}{\mathcal{C}(x^j)} \sum_{l=1}^{j} \sum_{k=1}^{K}f(x_i^j)g(x^l(p_i+\Delta p_{ik})),
% \end{equation}
% where $f$ is the function to generate attention weights from the query feature. The structure of CLA is illustrated in Figure~\ref{fig:CLA}. Similar to the design for non-local attention module in~\cite{wang2018non}, when deploying CLA in CNN backbones, we also wrap it into a non-local block with residual connection.

With such cross-layer design, each query feature from the input feature map refers to only a fixed number, $K$, of keys from each previous layer. However, query features at different spatial positions may have different preferences on keys sampled from different layers. The restoration process at different spatial positions may vary significantly due to the diversity of textures in an image, especially for image restoration tasks. As a result, the number of most semantically similar keys at each layer may not be the same across different layers.

% Besides, when deploying CLA in CNN backbones for image restoration, we find that increasing the number of inserted CLA modules would not constantly improve the performance while incurring much higher inference cost. This observation motivates us to search for optimal insert positions of ACLA modules in neural networks to reduce the inference cost while maintaining competitive performance.

%the optimal positions where ACLA should be inserted into the neural backbone

%

\begin{figure*}[!htbp]
\begin{center}
  \includegraphics[width=1.0\linewidth]{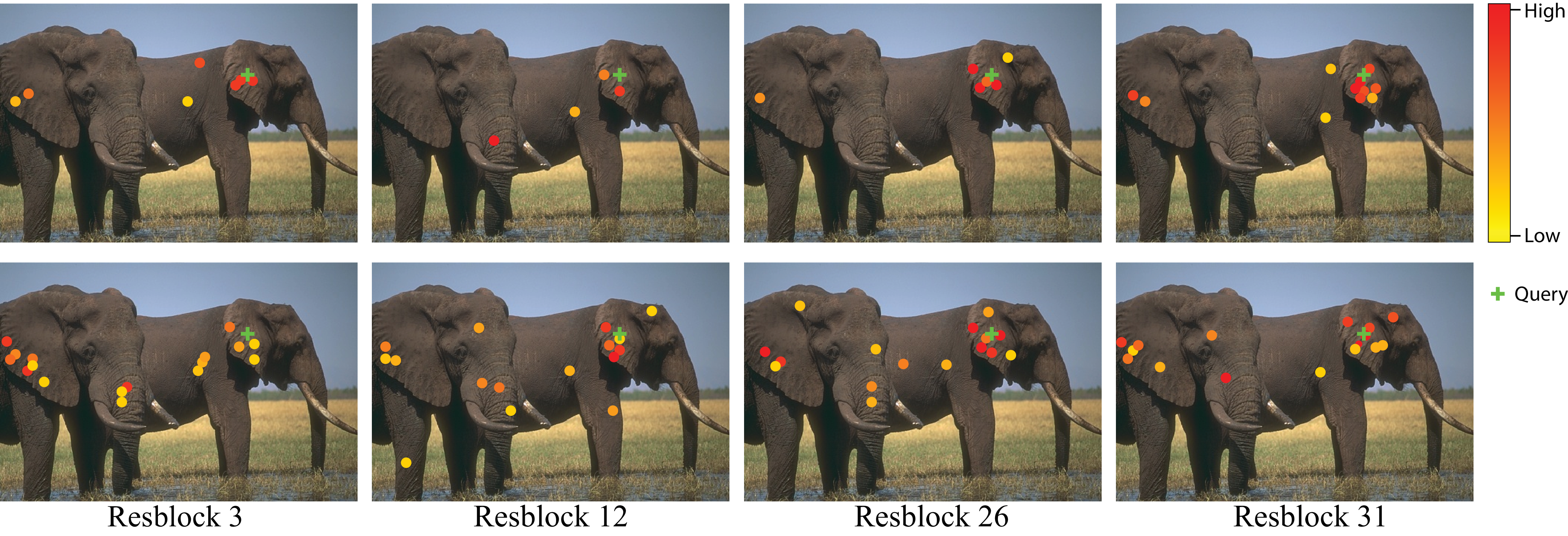}
\end{center}
\caption{Visualization of selected keys by ACLA for a query feature from the 31st resblock.~ The first row shows the positions of the keys selected by ACLA with $K=16$. For comparison, the positions of keys with top-$16$ attention weights following the CLNL formulation in Equation~(\ref{eq:cl_nl}) is displayed in the second row. From left to right are the sampled key positions from the 3rd, 12th, 26th, and 31st resblock. The query feature is shown as a green cross marker. Each sampled key feature is marked as a circle whose color indicates its attention weight. It can be observed that ACLA adaptively selects semantically similar key features for the query feature, while its vanilla counterpart lacks such capability. More visualization results and analysis can be found in Section~\ref{subsec:visualization}.}
\label{fig:key_sample_vis}
\end{figure*}
% \begin{figure}[!htbp]
% \begin{center}
% % \fbox{\rule{0pt}{2in} \rule{0.9\linewidth}{0pt}}
%   \includegraphics[width=1.4\columnwidth]{figs/gating_mask.eps}
% \end{center}
% \vspace{-3mm}

% \caption{Illustration of adaptive key selection in Adaptive Cross-Layer Attention (ACLA). The mask unit $\mathcal{M}$ generates a soft gating mask on the fixed number sampled keys, after which the Gumbel-Softmax turns the soft gating mask into a hard gating mask. The hard gating mask is applied on the features of the sampled keys and their corresponding attention weights before aggregation.}
% \label{fig:adaptive cross-layer_attention}
% \end{figure}

To achieve adaptive key selection in the cross-layer attention, we propose Adaptive Cross-Layer Attention (ACLA). Specifically, for each query feature, we dynamically search for the keys sampled from previous layers with ACLA. Besides, when deploying ACLA in CNN backbones, a neural architecture search method is used to search for the insert positions of ACLA. An objective based on the inference cost of inserted ACLA modules is used to supervise the search procedure.

To search for the informative sampled keys for a query feature from its previous layers, we apply a hard gating mask on the keys sampled from previous layers as
\begin{equation} \label{eq:final_csnl_search}
    y_i^{j} = \frac{1}{\mathcal{C}(x^j)} \sum_{l=1}^{j} \sum_{k=1}^{K}m_{i,k}^{j,l}  f(x_i^j)g(x^l(p_i+\Delta p_{ik})),
\end{equation}
where $m_{i,k}^{j,l}$ is a binary hard gating mask for the $k$-th sampled key from $x^l$ for query feature $x_i^j$, whose value is either $1$ or $0$. Compared to vanilla cross-layer attention, ACLA is more selective when aggregating key features to obtain the output feature. At layer $l$, it is expected that the most semantically similar keys, which correspond to nonzero $m_{i,k}^{j,l}$, are used to generate the output feature.

To optimize the hard gating mask with gradient descent, we relax the hard gating mask into the continuous domain with the simplified binary Gumbel-Softmax \cite{verelst2020dynamic}. Thus, the hard gating mask $m_{i,k}^{j,l}$ can be approximated by
\begin{equation}\label{eq:hard_mask}
\hat m_{i,k}^{j,l} = \sigma \Bigl( \frac{ \beta_{i,k}^{j,l} + \epsilon_{i,k,1}^{j,l} - \epsilon_{i,k,2}^{j,l} }{\tau} \Bigr),
\end{equation}
where $\hat m_{i,k}^{j,l}$ is an approximation of the hard gating mask $m_{i,k}^{j,l}$ in continuous domain. $\beta_{i,k}^{j,l}$ is the sampling parameter. $\epsilon_{i,k,1}^{j,l}, \epsilon_{i,k,2}^{j,l}$ are Gumbel noise for the approximation. $\tau$ is the temperature, and $\sigma$ is the Sigmoid function. During the training, the straight-through estimator from \cite{Bengio2013EstimatingComputation,verelst2020dynamic} is used for $m_{i,k}^{j,l}$. In the forward pass, the hard gating mask is computed by
\begin{equation}
\label{eq:straight_through}
m_{i,k}^{j,l} =
\begin{cases}
1           &\hat m_{i,k}^{j,l} >    0.5, \\
0           &\hat m_{i,k}^{j,l} \leq 0.5.
\end{cases}
\end{equation}
In the backward pass, we set $m_{i,k}^{j,l} = \hat m_{i,k}^{j,l}$ to enable the regular factional gradient used in stochastic gradient descent.

The sampling parameter $\beta_{i,k}^{j,l}$ in Equation (\ref{eq:hard_mask}) can be regarded as a soft gating mask, which is used to generate the hard gating mask. To achieve input-dependent key selection, a mask unit $\mathcal{M}$ is used to generate the soft gating mask $\beta$ from the features of the sampled keys as
\begin{equation} \label{eq:mask_unit}
\beta_{i,k}^{j,l} = \mathcal{M}(x^l(p_i+\Delta p_{ik})).
\end{equation}
Following the design in \cite{verelst2020dynamic}, a $1\times 1$ convolution layer is used as the mask unit $\mathcal{M}$ in our model. Gumbel noise $\epsilon_{i, l, 1}^k$ and $\epsilon_{i, l, 2}^k$ are set to $0$ during inference. With such a design, we are able to generate a soft gating mask from features of sampled keys and turn it into a hard gating mask to achieve the search for sampled keys based on the input. The overall framework of adaptive key selection in an ACLA module is illustrated in Figure~\ref{fig:ACLA}.
% We would like to emphasize that while both CLA and ACLA have a parameter $K$, the meaning of this parameter is significantly different for each method. While CLA samples a fixed number, $K$, of keys at each layer, ACLA would only sample at most $K$ keys at each layer.
% The number of sampled keys at layer by ACLA is determined by optimization of the neural network with ACLA modules, which includes the searching process for the informative sampled keys at each layer.

To demonstrate the effectiveness of adaptive key selection in ACLA, we compare the keys selected by ACLA and those selected by vanilla Cross-Layer Non-Local (CLNL) at different layers for a query feature in Figure~\ref{fig:key_sample_vis}. It can be observed that semantically similar keys are selected by ACLA for the query feature.
\begin{figure*}[!ht]
\begin{center}
  \includegraphics[width=0.8\linewidth]{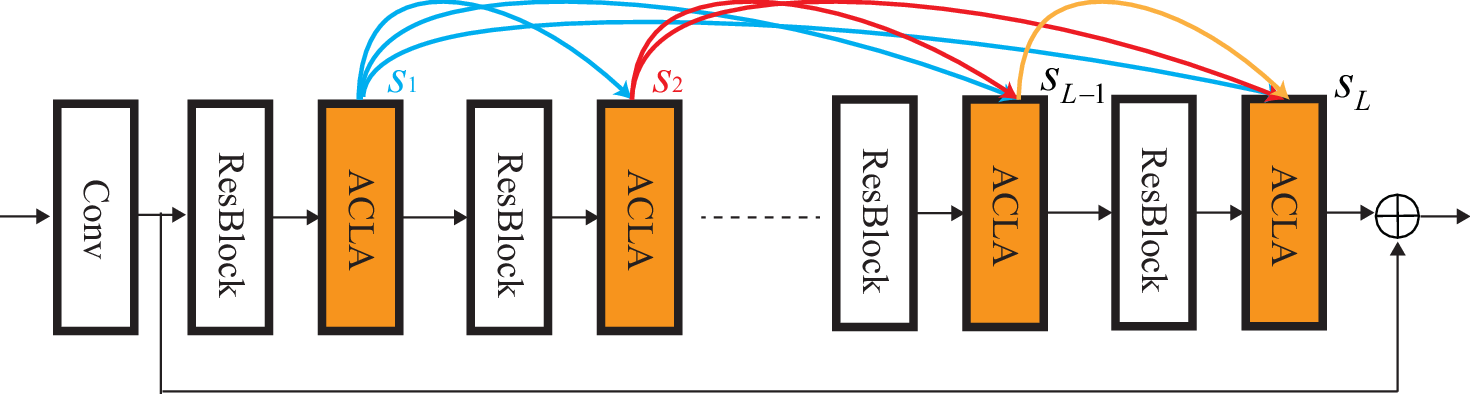}
\end{center}
\caption{Illustration of the search for insert positions in ACLA. Except for image super-resolution where the first Conv block is an upscaling block that increases the image resolution, the first Conv block maintains the resolution of the input for the other image restoration tasks.}
\label{fig:search_space}
\end{figure*}

\subsection{Insert Positions for ACLA}
\label{sec:ACLA-insert-position}

\textbf{Insert Positions for ACLA.~} As demonstrated in Section~\ref{sec:ablation-two-adaptive-designs}, the positions where ACLA modules are inserted into the neural backbone have a considerable effect on the final performance. In order to decide the insert positions of ACLA modules in a neural backbone, we propose the following search method.
We first densely insert ACLA after each layer of the CNN backbone as shown in Figure~\ref{fig:search_space} to build the supernet. Similar to the gating formulation in ACLA, we define a hard decision parameter $s_l \in \{0,1\}$ for the $l$-th inserted ACLA in the supernet. $s_l=1$ indicates that an ACLA module is inserted after the $l$-th layer, and $s_l=0$ otherwise. As a result, the output of ACLA in the supernet can be expressed as
\begin{equation} \label{eq:masked_csnl_search}
    y_i^{j} = \frac{1}{\mathcal{C}(x^j)} \sum_{l=1}^{j} s_l \sum_{k=1}^{K}m_{i,k}^{j,l}  f(x_i^j)g(x^l(p_i+\Delta p_{ik})).
\end{equation}
It can be observed from (\ref{eq:masked_csnl_search}) that the output of ACLA is the aggregation of features of adaptive keys at previous layers selected by $\left\{s_l\right\}$. It is worthwhile to mention that the search for insertion positions enables automatic search for referred layers for each query feature. In particular, the referred layers of a query $x_i^j$ are those of index $l$ with the decision parameter $s_l = 1$ and at least one nonzero mask in
the binary hard gating mask
$\{m_{i,k}^{j,l}\}$.

The simplified binary Gumbel-Softmax \cite{verelst2020dynamic} is used here to approximate the hard decision parameter $s_l$ by
\begin{equation}
\hat{s}_l = \sigma \Bigl( \frac{ \alpha_l + \epsilon_{1}^j - \epsilon_{2}^j }{\tau} \Bigr),
\end{equation}
with sampling parameter $\alpha_l$, Gumbel noise $\epsilon$, and temperature $\tau$.
Different from the input-dependent design of the gating mask in ACLA, here we directly replace $s_l$ with its continuous approximation $\hat{s}_l$. $\alpha_l$ here can be regarded as architecture parameters and can be directly optimized by stochastic gradient descent (SGD) during the search process. By gradually decreasing the temperature $\tau$, $\alpha_l$ will be optimized such that $s_l$ will approach $1$ or $0$.

\textbf{Search Procedure.~} To render a compact and efficient neural network with ACLA modules, we need to optimize both the accuracy of a neural network and the inference cost (FLOPs) of the ACLA modules inserted into that neural network. Therefore, the inference cost of the ACLA modules inserted needs to be estimated during the search phase. Following the formulation of the ACLA in the supernet, the inference cost of the ACLA inserted after the $j$-th residual block as
\begin{equation} \label{eq:cost_csnl}
    \mathtt{cost}_j =  \sum_{l=1}^{j} s_l \sum_{k=1}^{K} (2 m_{i,k}^{j,l} NC^2+2NC^2+6KNC),
\end{equation}
where $N$ is the number of spatial positions, $C$ is the number of channels, $K$ is the maximal number of sampled keys. $2m_{j, l}^k NC^2$ is the FLOPs for the convolution on generating the gating masks. $2NC^2+6KNC$ is the FLOPs for generating the attention weights and 2D offsets. Then we obtain the inference cost of all inserted ACLA modules as
\begin{equation} \label{eq:cost_all}
    \mathtt{cost} =  \sum_{j=1}^{L} s_j \mathtt{cost}_j.
\end{equation}
As mentioned before, due to relaxation to continuous problems, we search for the architecture of ACLA, which is comprised of sampled keys at each layer and the insert positions of ACLA modules, by updating the architecture parameters using SGD. The architecture parameters of ACLA are $\alpha = \{\alpha_j\}_{j=1}^L$, where $L$ is the number of layers in the neural network with ACLA. To supervise the search process, we design a loss function with cost-based regularization to achieve multi-objective optimization:
\begin{equation}
    \label{eq: loss}
    \mathcal{L}(w, \alpha) = \mathcal{L}_{MSE} + \lambda \log\mathtt{cost},
\end{equation}
where $\lambda$ is the hyper-parameters that control the magnitude of the cost term.

We find that at the beginning of the search process, ACLA modules inserted at shallow layers are more likely to be maintained. Similar problems has been observed by previous NAS works \cite{fang2020densely}. To solve this problem, we follow DCNAS \cite{fang2020densely} and split our search procedure into two stages. In the first stage, we only optimize the parameters of the network for enough epochs to get network weights sufficiently trained. In the second stage, we activate the architecture optimization. We alternatively optimize the network weights by descending $\nabla_w \mathcal{L}_{train}(w, \alpha)$ on the training set, and optimize the architecture parameters by descending $\nabla_{\alpha} \mathcal{L}_{val}(w, \alpha)$ on the validation set. When the search procedure terminates, we derive the insert positions based on the architecture parameters $\alpha$.

\textbf{Differences from Deformable DETR \cite{zhu2020deformable}.}
The proposed ACLA is significantly different from Deformable DETR \cite{zhu2020deformable}. Deformable DETR proposes a sparse attention module where each query only attends to a small and fixed set of sampled keys in the input feature map by learning their 2D offsets from the query point. Then, it aggregates the key features selected with the query features.
As discussed earlier, Deformable DETR suffers from lack of keys across different layers and lack of flexibility in sampled keys across different layers. It is demonstrated by the ablation studies in Section~\ref{sec:ablation-two-adaptive-designs} and Section~\ref{sec:ablation-referred-layers} that referring to previous layers and adaptive key selection improve the performance of attention mechanism compared to baselines without these characteristics.

In contrast with Deformable DETR, each query in ACLA attends to keys from previous layers. Furthermore, ACLA learns hard gating masks for the keys selected from the previous layers. Multiplying the gating masks by the keys selected from the previous layers, each query dynamically selects an adaptive number of keys from each previous layer to attend to. The hard gating masks are obtained by applying Gumbel-Softmax to soft gating masks learned from the query feature in the continuous domain. In addition, the optimal insert positions of ACLA modules are decided using a differentiable neural architecture search algorithm. A set of architecture parameters defined for each layer in the network are learned by optimizing both the MSE loss and the inference cost (FLOPs) of the network. As a result, each query in ACLA attends to an adaptive number of keys from feature maps at selected previous layers.

\begin{table*}[htb]
    \centering
    \caption{Quantitative results on benchmark datasets for single image super-resolution. The performance of the best baseline is underlined.}
    \vspace{-3mm}
        \label{table:SR_results}
        \resizebox{1.0\textwidth}{!}{
\begin{tabular}{|c|c|c|c|c|c|c|c|c|c|c|c|c|}
\hline
\multirow{2}{*}{Method} & \multirow{2}{*}{Scale} & \multirow{2}{*}{Params(M)} & \multicolumn{2}{c|}{Set5}           & \multicolumn{2}{c|}{Set14}         & \multicolumn{2}{c|}{B100}           & \multicolumn{2}{c|}{Urban100}       & \multicolumn{2}{c|}{Manga109}       \\
\cline{4-13}
                        &                        &                            & PSNR              & SSIM            & PSNR             & SSIM            & PSNR              & SSIM            & PSNR              & SSIM            & PSNR             & SSIM             \\
\hline
Bicubic                 & $\times$2              & -                          & 33.66             & 0.9299          & 30.24            & 0.8688          & 29.56             & 0.8431          & 26.88             & 0.8403          & 30.80            & 0.9339           \\
SRCNN                   & $\times$2              & 0.244                      & 36.66             & 0.9542          & 32.45            & 0.9067          & 31.36             & 0.8879          & 29.50             & 0.8946          & 35.60            & 0.9663           \\
VDSR                    & $\times$2              & 0.672                      & 37.53             & 0.9590          & 33.05            & 0.9130          & 31.90             & 0.8960          & 30.77             & 0.9140          & 37.22            & 0.9750           \\
MemNet                  & $\times$2              & 0.677                      & 37.78             & 0.9597          & 33.28            & 0.9142          & 32.08             & 0.8978          & 31.31             & 0.9195          & 37.72            & 0.9740           \\
SRMDNF                  & $\times$2              & 5.69                       & 37.79             & 0.9601          & 33.32            & 0.9159          & 32.05             & 0.8985          & 31.33             & 0.9204          & 38.07            & 0.9761           \\
RDN                     & $\times$2              & 22.6                       & 38.24             & 0.9614          & 34.01            & 0.9212          & 32.34             & 0.9017          & 32.89             & 0.9353          & 39.18            & 0.9780           \\
SAN                     & $\times$2              & 16.7                       & 38.31             & 0.9620          & 34.07            & 0.9213          & 32.42             & 0.9028          & 33.10             & 0.9370          & 39.32            & 0.9792           \\
HAN                     & $\times$2              & 17.3                       & 38.27             & 0.9614          & \underline{34.16}    & 0.9217  & 32.41             & {0.9027}  & 33.35             & 0.9385          & 39.46            & 0.9787           \\
SwinIR                    & $\times$2               & 11.8                       & {38.35}     & {0.9620}  & 34.14            & {0.9227}          & \underline{32.42}     & \underline{0.9030}  & {33.40}     & {0.9393}  & {39.59}    & {0.9790}  \\
HAT                    & $\times$2               & 24.8                       & \underline{38.34}     & \underline{0.9621}  & 34.11            & \underline{0.9232}          & {32.40}     & {0.9028}  & \underline{33.52}     & \underline{0.9400}  & \underline{39.60}    & \underline{0.9792}  \\
NLSN                    & $\times$2               & 44.3                       & \underline{38.34}     & {0.9618}  & 34.08            & {0.9231}          & {32.43}     & {0.9027}  & {33.42}     & {0.9394}  & {39.59}    & {0.9789}  \\
\hline
EDSR                    & $\times$2              & 40.7                       & 38.11             & 0.9602          & 33.92            & 0.9195          & 32.32             & 0.9013          & 32.93             & 0.9351          & 39.10            & 0.9773           \\
EDSR+NL               & $\times$2              & 43.6                       & 38.15             & 0.9606          & 34.00            & 0.9203          & 32.37             & 0.9021          & 33.05             & 0.9360          & 39.21            & 0.9778           \\
ACLA               & $\times$2              & 42.3                       & \textbf{38.39}    & \textbf{0.9623} & \textbf{34.24}   & \textbf{0.9234} & \textbf{32.55}    & \textbf{0.9038} & \textbf{33.56}    & \textbf{0.9403} & \textbf{39.77}   & \textbf{0.9789}  \\
\hline
p-value        & $\times$2              & -                          & 0.0021 & -               & 0.0010  & -               & 1.99e-12 & -               & 3.74e-10 & -               & 1.53e-6  & -                \\
\hline
\hline
Bicubic                 & $\times$3              & -                          & 30.39             & 0.8682          & 27.55            & 0.7742          & 27.21             & 0.7385          & 24.46             & 0.7349          & 26.95            & 0.8556           \\
SRCNN                   & $\times$3              & 0.244                      & 32.75             & 0.9090          & 29.30            & 0.8215          & 28.41             & 0.7863          & 26.24             & 0.7989          & 30.48            & 0.9117           \\
VDSR                    & $\times$3              & 0.672                      & 33.67             & 0.9210          & 29.78            & 0.8320          & 28.83             & 0.7990          & 27.14             & 0.8290          & 32.01            & 0.9340           \\
MemNet                  & $\times$3              & 0.677                      & 34.09             & 0.9248          & 30.00            & 0.8350          & 28.96             & 0.8001          & 27.56             & 0.8376          & 32.51            & 0.9369           \\
SRMDNF                  & $\times$3              & 5.69                       & 34.12             & 0.9254          & 30.04            & 0.8382          & 28.97             & 0.8025          & 27.57             & 0.8398          & 33.00            & 0.9403           \\
RDN                     & $\times$3              & 22.6                       & 34.71             & 0.9296          & 30.57            & 0.8468          & 29.26             & 0.8093          & 28.80             & 0.8653          & 34.13            & 0.9484           \\
SAN                     & $\times$3              & 16.7                       & 34.75             & 0.9300          & 30.59            & 0.8476          & 29.33             & 0.8112          & 28.93             & 0.8671          & 34.30            & 0.9494           \\
HAN                     & $\times$3              & 17.3                       & 34.75             & 0.9299          & 30.67            & 0.8483          & 29.32             & 0.8110          & 29.10             & 0.8705          & 34.48            & 0.9500           \\
SwinIR                    & $\times$3               & 11.8                       & \underline{34.86}             & \underline{0.9310}         &30.70       &0.8484   & 29.31             &  0.8115 &  29.24    & 0.8726  &   34.56      & 0.9507   \\
HAT                    & $\times$3               & 24.8                       & 34.84    &0.9305   &  \underline{30.71}           & \underline{0.8485}         &   29.31  & 0.8116 &  \underline{29.28}   &\underline{0.8728}   &  \underline{34.57}  &\underline{0.9509}   \\
NLSN                    & $\times$3               & 44.3                       & {34.85}     & {0.9306}  & {30.70}    & \underline{0.8485}  & \underline{29.34}     & \underline{0.8117}  & {29.25}     & {0.8726}  & \underline{34.57}    & {0.9508}   \\
\hline
EDSR                    & $\times$3              & 40.7                       & 34.65             & 0.9280          & 30.52            & 0.8462          & 29.25             & 0.8093          & 28.80             & 0.8653          & 34.17            & 0.9476           \\
EDSR+NL                 & $\times$3              & 43.6                       & 34.70             & 0.9291          & 30.57            & 0.8470          & 29.26             & 0.8102          & 28.87             & 0.8670          & 34.22            & 0.9484           \\
ACLA               & $\times$3              & 42.3                       & \textbf{34.91}    & \textbf{0.9312} & \textbf{30.80}   & \textbf{0.8494} & \textbf{29.43}    & \textbf{0.8127} & \textbf{29.40}    & \textbf{0.8734} & \textbf{34.71}   & \textbf{0.9516}  \\
\hline
p-value       & $\times$3              & -                          & 0.0015   & -               & 0.0003 & -               & 8.98e-8 & -               & 2.44e-9 & -               & 4.64e-6  & -                \\
\hline
\hline
Bicubic                 & $\times$4              & -                          & 28.42             & 0.8104          & 26.00            & 0.7027          & 25.96             & 0.6675          & 23.14             & 0.6577          & 24.89            & 0.7866           \\
SRCNN                   & $\times$4              & 0.244                      & 30.48             & 0.8628          & 27.50            & 0.7513          & 26.90             & 0.7101          & 24.52             & 0.7221          & 27.58            & 0.8555           \\
VDSR                    & $\times$4              & 0.672                      & 31.35             & 0.8830          & 28.02            & 0.7680          & 27.29             & 0.0726          & 25.18             & 0.7540          & 28.83            & 0.8870           \\
MemNet                  & $\times$4              & 0.677                      & 31.74             & 0.8893          & 28.26            & 0.7723          & 27.40             & 0.7281          & 25.50             & 0.7630          & 29.42            & 0.8942           \\
SRMDNF                  & $\times$4              & 5.69                       & 31.96             & 0.8925          & 28.35            & 0.7787          & 27.49             & 0.7337          & 25.68             & 0.7731          & 30.09            & 0.9024           \\
RDN                     & $\times$4              & 22.6                       & 32.47             & 0.8990          & 28.81            & 0.7871          & 27.72             & 0.7419          & 26.61             & 0.8028          & 31.00            & 0.9151           \\
SAN                     & $\times$4              & 16.7                       & 32.64             & 0.9003          & 28.92            & 0.7888          & 27.78             & 0.7436          & 26.79             & 0.8068          & 31.18            & 0.9169           \\
HAN                     & $\times$4              & 17.3                       & {32.64}     & {0.9002}  & \underline{28.90}    & 0.7890          & \underline{27.80}     & 0.7442          & 26.85             & 0.8094          & \underline{31.42}    & 0.9177           \\
SwinIR                    & $\times$4               & 11.8    &     \underline{32.65}       & 0.9014         & 28.89      &0.7890   &  27.78            & 0.7443  & 26.95     & \underline{0.8150} & 31.33      & 0.9180   \\
HAT                    & $\times$4               & 24.8                       & \underline{32.65}    & \underline{0.9015}  &  28.86           & \underline{0.7892}         & 27.76    &0.7441  &  \underline{26.97}   &0.8113   & 31.30   &  0.9183 \\
NLSN                    & $\times$4               & 44.3                       & 32.59             & 0.9000          & 28.87            & {0.7891}  & 27.78             & \underline{0.7444}  & {26.96}     & {0.8159}  & 31.27            & \underline{0.9184}   \\
\hline
EDSR                    & $\times$4              & 40.7                       & 32.46             & 0.8968          & 28.80            & 0.7876          & 27.71             & 0.7420          & 26.64             & 0.8033          & 31.02            & 0.9148           \\
EDSR+NL                 & $\times$4              & 43.6                       & 32.53             & 0.8994          & 28.82            & 0.7877          & 27.74             & 0.7430          & 26.71             & 0.8069          & 31.19            & 0.9154           \\
ACLA               & $\times$4              & 42.3                       & \textbf{32.70}    & \textbf{0.9020} & \textbf{28.98}   & \textbf{0.7910} & \textbf{27.86}    & \textbf{0.7460} & \textbf{27.12}    & \textbf{0.8170} & \textbf{31.53}   & \textbf{0.9215}  \\
\hline
p-value        & $\times$4              & -                          & 0.0012   & -               & 0.0007  & -               & 0.0005  & -               & 4.57e-12 & -               & 3.62e-9 & -                \\
\hline
\end{tabular}

        }
\end{table*}

\begin{table*}[!htbph]
%\scriptsize
%\footnotesize
%\small
%\normalsize
\center
\begin{center}
\caption{Quantitative results on benchmark datasets for single image denoising}
\label{tab:results_psnr_denoise_rgb}
\vspace{-3mm}
\resizebox{1.0\textwidth}{!}{
\begin{tabular}{|c|c|c|c|c|c|c|c|c|c|c|c|c|c|}
\hline
\multirow{2}{*}{Method} & \multirow{2}{*}{Params (M)} & \multicolumn{4}{c|}{KCLDAk24}                                                & \multicolumn{4}{c|}{BSD68}                                                 & \multicolumn{4}{c|}{Urban100}                                                \\
\cline{3-14}
                        &                             & 10                & 30               & 50                & 70                & 10                & 30               & 50                & 70              & 10               & 30                & 50                & 70                \\
\hline
MemNet                  & 0.677                       & N/A               & 29.67            & 27.65             & 26.40             & N/A               & 28.39            & 26.33             & 25.08           & N/A              & 28.93             & 26.53             & 24.93             \\
DnCNN                   & 0.672                       & 36.98             & 31.39            & 29.16             & 27.64             & 36.31             & 30.40            & 28.01             & 26.56           & 36.21            & 30.28             & 28.16             & 26.17             \\
RNAN                    & 7.41                       & 37.24             & 31.86            & 29.58             & 28.16             & 36.43             & 30.63            & 28.27             & 26.83           & 36.59            & 31.50             & 29.08             & 27.45             \\
PANet                   & 5.96                       & {37.35}             & {31.96}    & {29.65}             & {28.20}             & {36.50}             & {30.70}            & {28.33}             & \underline{26.89}           & {36.80}    & {31.87}     & {29.47}     & {27.87}             \\
SwinIR                      &11.8                       & 37.38             & 31.97            &\underline{29.67}             & 28.20            & 36.50             & 30.71            &  \underline{28.35}          &  26.87        & 36.84            & 31.88             &   29.48         & 27.89         \\
SCUNet                      &10.8                       & \underline{37.41}             & \underline{31.99}            &  29.65           & \underline{28.23}            & \underline{36.52}             & 30.71            &  \underline{28.35}           & 26.85          & \underline{36.87}            & \underline{31.91}             & 29.48             &  \underline{27.90}            \\
Restormer                      &15.8                       & 37.40             & 31.96            &  \underline{29.67}           & 28.20           & 36.50             & \underline{30.73}            &  28.33           & 26.87          & 36.85            & 31.90             & \underline{29.51}             &  27.89            \\
\hline
Baseline                & 5.43                       & 37.21             & 31.85            & 29.60             & 28.15             & 36.34             & 30.60            & 28.28             & 26.84           & 36.63            & 31.64             & 29.22             & 27.54             \\
NL                      & 6.14                       & 37.29             & 31.90            & 29.64             & 28.19             & 36.43             & 30.67            & 28.31             & 26.89           & 36.69            & 31.74             & 29.30             & 27.70             \\
% CLA                     & 5.896                       & {37.37}     & \textbf{31.97}   & {29.67}     & {28.23}     & {36.52}     & {30.74}    & {28.35}     & {26.91}   & 36.79            & 31.85             & 29.43             & {27.88}     \\
ACLA                    & 5.91                       & \textbf{37.52 }   & \textbf{32.10}   & \textbf{29.78}    & \textbf{28.33 }   & \textbf{36.65}    & \textbf{30.83}   & \textbf{28.47 }   & \textbf{26.99 } & \textbf{36.97}   & \textbf{31.99}    & \textbf{29.63 }   & \textbf{ 27.99}   \\
\hline
p-value        & -                 & 3.95e-13 & 6.75e-9 & 4.23e-12 & 3.87e-12 & 8.19e-11 & 7.54e-9 & 5.29e-12 & 9.31e-11  & 7.97e-10 & 2.50e-10 & 2.97e-11 & 1.77e-11  \\
\hline
\end{tabular}
}
\end{center}
%\vspace{-5mm}
\end{table*}
\section{Experiments}
\label{sec:experiments}
In this section, we evaluate the performance of ACLA on image restoration tasks, including single image super-resolution, image denoising, image compression artifacts reduction, and image demosaicing. In the implementation, ACLA is deployed on the commonly used neural network model, EDSR~\cite{EDSR}, for all image restoration tasks. Comparisons with competing methods demonstrate the effectiveness of ACLA. In addition, we perform t-test between ACLA and the current SOTA methods to show the statistical significance of improvement for each task.

\subsection{Implementation Details}
\label{sec:settings}
We use DIV2K~\cite{timofte2017ntire} as the training set and EDSR \citep{lim2017enhanced} as the neural backbones for different image restoration tasks. Following previous works\cite{EDSR, mei2021image, mei2020pyramid}, we use EDSR with 32 residual blocks as the backbones for image super-resolution and  EDSR with 16 residual blocks as the backbones for image denoising, image compression artifacts reduction, and image demosaicing. In our experiments, ACLA modules are inserted between different residual blocks. DIV2K consists of $800$ images for training and $100$ images for validation. We follow the training settings in previous works \cite{dai2019second,zhang2018residual, EDSR, niu2020single} for fair comparisons. We augment the training images by randomly rotating $90$\degree, $180$\degree, $270$\degree, and horizontally flipping. In each mini-batch, $16$ low-quality patches with size $48 \times 48$ are provided as inputs. ADAM optimizer is used for both the search phase and training phase. Default values of $\beta_{1}$ and $\beta_{2}$ are set to $0.9$ and $0.999$ respectively, and we set $\epsilon=10^{-8}$. In the search phase, the learning rate is initialized as $10^{-4}$, and the cosine learning rate schedule is used. The search process takes 600 epochs. The first stage of the search takes 300 epochs, and the second stage takes the remaining 300 epochs. In the training phase, the learning rate is initialized as $10^{-4}$ and the cosine learning rate schedule is used to decay the learning rate to $5\times 10^{-6}$ in $800$ epochs.
% In our experiments with CLA, we manually insert four CLA modules evenly in the CNN backbones. The number of selected keys $K$ is set to $8$ for all CLA modules by default.
For all ACLA modules, the maximum number of selected keys, which is also denoted by $K$, is initialized as $16$. Before the search, we perform a cross-validation on $20\%$ of the training data to decide the value of $\lambda$. Another $10\%$ of training data is held for evaluation in the cross-validation process. The hyper-parameter $\lambda$ is selected from a candidate set $\{0.1, 0.15, 0.2, 0.25, 0.3, 0.35, 0.4\}$. The selected $\lambda$ for different tasks are summarized in Table \ref{table:lamda_setting}.

\subsection{Single Image Super-Resolution}
\begin{figure*}[htb]
\begin{center}
  \includegraphics[width=0.75\textwidth]{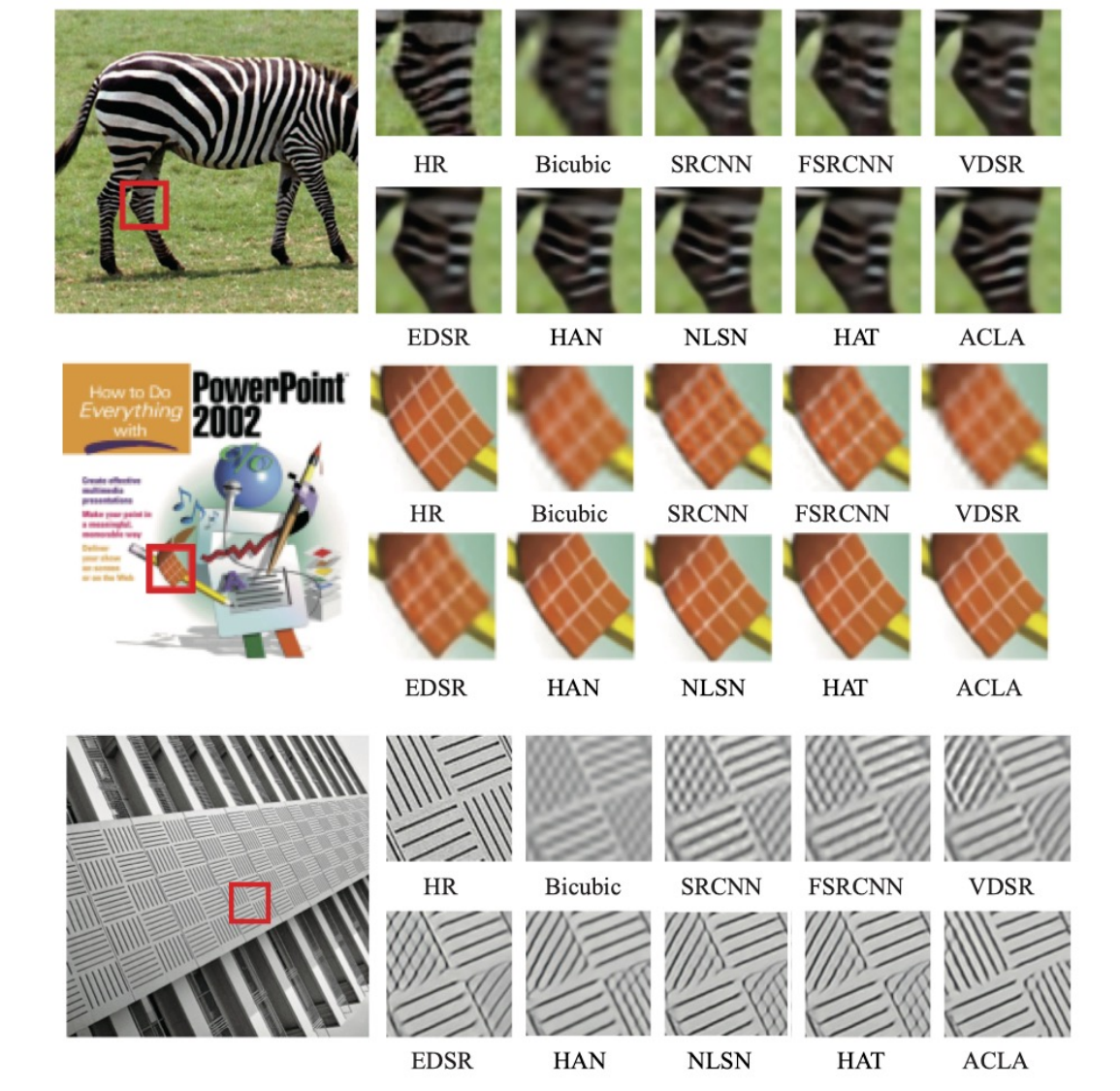}
\end{center}
\caption{Visual comparison for 4$\times$ SR with BI degradation model.}
\vspace{-2mm}
\label{fig:visual_SR}
\end{figure*}

For single image super-resolution, we evaluate ACLA on top of the widely used super-resolution backbone EDSR~\cite{lim2017enhanced}. The LR images are obtained by the bicubic downsampling of HR images. All the methods are evaluated on five standard datasets, Set5~\cite{bevilacqua2012low}, Set14~\cite{zeyde2010single}, B100~\cite{martin2001database}, Urban100~\cite{huang2015single}, and Manga109~\cite{matsui2017sketch}. The reconstructed results by our model are converted to YCbCr space. PSNR and SSIM in the luminance channel are calculated in our experiments. We compare our method with six baseline methods, SRCNN \cite{dong2015image}, VDSR \cite{kim2016accurate}, MemNet \cite{tai2017memnet}, SRMDNF \cite{zhang2018learning}, RDN \cite{zhang2018residual}, SAN \cite{SAN}, HAN \cite{niu2020single}, NLSN \cite{mei2021image}, SwinIR\cite{liang2021swinir}, and HAT \cite{chen2022activating}. Note that the results of SwinIR and HAT reported in \cite{liang2021swinir, chen2022activating} are obtained by models trained on DIV2K \cite{timofte2017ntire} and Flick2K \cite{EDSR}. In our experiments, we train SwinIR and HAT on DIV2K with the same settings as ACLA for fair comparisons. The quantitative results are shown in Table~\ref{table:SR_results}. The visual comparisons between ACLA and previous baselines are shown in Figure~\ref{fig:visual_SR}. Our method greatly improves the performance of EDSR on all benchmarks with all upsampling scales. \textbf{In particular, the improvements of PSNR over the top baselines NLSN/HAT for $2\times$, $3 \times$, and $4 \times$ image super-resolution, averaged over all the benchmarks, are 0.118 dB, 0.110 dB, and 0.112 dB respectively}.
To verify that such improvement is statistically significant and out of the range of error, we train ACLA and the top baselines, NLSN and HAT, on different super-resolution scales ten times with different seeds for random initialization of the networks. The mean and standard deviation of different runs are shown in Table~\ref{table:SR_q}. Then we perform t-test between the results of ACLA and the best among NLSN and HAT on each benchmark dataset with all the super-resolution scales. The largest p-value among all the datasets and super-resolution scales is $0.0021 \ll 0.05$  on Set5 for $2\times$ super-resolution, suggesting that the improvement of ACLA over NLSN is statistically significant.

\begin{table*}[bht]
    \centering
    \caption{Quantitative results on benchmark datasets for image compression artifacts reduction}
    \vspace{-3mm}
        \label{table:CAR_results}
        %\vspace{-0.7em}
        \resizebox{0.7\textwidth}{!}{
\begin{tabular}{|c|c|c|c|c|c|c|c|c|c|}
\hline
\multirow{2}{*}{Method} & \multirow{2}{*}{Params (M)} & \multicolumn{4}{c|}{LIVE1}                                                & \multicolumn{4}{c|}{Classic5}                                           \\
\cline{3-10}
                        &                             & 10                & 20              & 30              & 40                & 10               & 20              & 30              & 40               \\
\hline
JPEG                    & -                           & 27.77             & 30.07           & 31.41           & 32.35             & 27.82            & 30.12           & 31.48           & 32.43            \\
DnCNN                   & 0.672                       & 29.19             & 31.59           & 32.98           & 33.96             & 29.40            & 31.63           & 32.91           & 33.77            \\
RNAN                    & 7.41                       & 29.63             & 32.03           & 33.45           & 34.47             & 29.96            & 32.11           & 33.38           & 34.27            \\
PANet                   & 5.96                       & {29.69}     & {32.10}           & {33.55}           & {34.55}             & {30.03}            & {32.36}           & {33.53}           & {34.38}            \\
SwinIR                   & 11.8                      & \underline{29.74}     & \underline{32.13}           & \underline{33.57}           & \underline{34.63}             & \underline{30.06}            & \underline{32.43}           & \underline{33.55}           & \underline{34.42}            \\
\hline
Baseline                & 5.43                       & 29.63             & 32.04           & 33.50           & 34.51             & 29.99            & 32.22           & 33.43           & 34.31            \\
NL                      & 6.14                       & 29.65             & 32.08           & 33.55           & 34.53             & 30.01            & 32.34           & 33.51           & {34.35}            \\
% CLA                     & 5.896                       & \textbf{29.73}    & {32.13}   & {33.57}   & {35.54}     & {30.05}    & {32.38}   & {33.55}   & {34.42}    \\
ACLA                    & 5.91                       & \textbf{29.83}    & \textbf{32.25}  & \textbf{33.68}  & \textbf{34.71}    & \textbf{30.20}   & \textbf{32.51}  & \textbf{33.67}  & \textbf{34.55}   \\
\hline
p-value        & -                  & 8.79e-12 & 0.0003 & 0.0002 & 5.75e-9 & 8.19e-8 & 0.0019 & 0.0024 & 0.0011  \\
\hline
\end{tabular}
        }

\end{table*}

\begin{table*}[thbp]\setlength{\tabcolsep}{10pt} \label{tab:results_psnr_demosaic_rgb}
\scriptsize
%\footnotesize
%\small
%\normalsize
\center
\begin{center}
%\vspace{-5mm}
\caption{Quantitative results on benchmark datasets for image demosaicing}
\label{tab:results_psnr_demosaic_rgb}
\vspace{-2mm}
%\begin{tabular*}{75.8mm}{@{\extracolsep{-0.99mm}}cccccccccc|c|c|c|c|c|c|c|c|c|c|c|c|c|c|c|c|c|c|}
\resizebox{0.9\textwidth}{!}{
\begin{tabular}{|c|c|c|c|c|c|c|c|c|c|}
\hline
\multirow{2}{*}{Method} & \multirow{2}{*}{Params(M)} & \multicolumn{2}{c|}{McMaster18}     & \multicolumn{2}{c|}{Kodak24}        & \multicolumn{2}{c|}{BSD68}         & \multicolumn{2}{c|}{Urban100}       \\
\cline{3-10}
                        &                            & PSNR              & SSIM            & PSNR              & SSIM            & PSNR             & SSIM            & PSNR             & SSIM             \\
\hline
Mosaiced                & -                          & 9.17              & 0.1674          & 8.56              & 0.0682          & 8.43             & 0.0850          & 7.48             & 0.1195           \\
IRCNN                   & 0.731                      & 37.47             & 0.9615          & 40.41             & 0.9807          & 39.96            & 0.9850          & 36.64            & 0.9743           \\
RNAN                    & 7.41                      & 39.71             & 0.9725          & 43.09             & 0.9902          & 42.50            & 0.9929          & 39.75            & 0.9848           \\
PANet                   & 5.96                      & \underline{40.00}             & \underline{0.9737}          & \underline{43.29}             & \underline{0.9905}          & \underline{42.86}            & \underline{0.9933}          & \underline{40.50}            & \underline{0.9854}   \\
\hline
Baseline                & 5.43                      & 39.81             & 0.9730          & 43.18             & 0.9903          & 42.66            & 0.9931          & 40.23            & 0.9852           \\
NL                      & 6.14                      & 39.90             & 0.9732          & 43.23             & 0.9903          & 42.79            & 0.9932          & 40.39            & 0.9853           \\
% CLA                     & 5.896                      & {40.03}     & {0.9739}  & {43.35}     & {0.9906}  & {42.88}    & {0.9934}  & {40.52}    & 0.9853           \\
ACLA                    & 5.91                      & \textbf{40.13}    & \textbf{0.9749} & \textbf{43.42}    & \textbf{0.9917} & \textbf{43.00}   & \textbf{0.9950} & \textbf{40.63}   & \textbf{0.9864}  \\
\hline
p-value        & -                 & 7.63e-10 & -      & 5.95e-12 & -      & 8.41e-11 & -      & 5.97e-10 & -       \\
\hline
\end{tabular}
 }
\end{center}
\end{table*}
\begin{table*}[!htbp]
\centering

    \caption{PSNR (mean/std) results comparison with p-value between ACLA and top baselines NLSN/HAT for single-image super-resolution}
    \vspace{-3mm}
        \label{table:SR_q}
        \resizebox{0.7\textwidth}{!}{
\begin{tabular}{|c|c|c|c|c|c|c|}
\hline
Methods & Scale      & Set 5           & Set 14          & B100            & Urban100        & Manga109         \\
\hline
NLSN     & $\times 2$ & 38.34 / 0.0033 & 34.08 / 0.0030 & 32.44 / 0.0041 & 33.42 / 0.0055 & 39.59 / 0.0056  \\
HAT     & $\times 2$ & 38.34 / 0.0033 & 34.11 / 0.0030 & 32.40 / 0.0041 & 33.52 / 0.0055 & 39.60 / 0.0056  \\
ACLA    & $\times 2$ & 38.39 /~0.0029 & 34.20 / 0.0048 & 32.55 / 0.0053 & 33.56 / 0.0070 & 39.77 / 0.0069  \\
\hline
p-value & $\times 2$ & 0.0021       & 0.0010          & 1.99e-12        & 3.74e-10        & 1.53e-6           \\
\hline
\hline
NLSN     & $\times 3$ & 34.85 / 0.0035 & 30.70 / 0.0028 & 29.34 / 0.0049 & 29.25 / 0.0052 & 34.57 / 0.0061  \\
HAT     & $\times 3$ & 34.84 / 0.0035 & 30.71 / 0.0028 & 29.31 / 0.0049 & 29.28 / 0.0052 & 34.57 / 0.0061  \\
ACLA    & $\times 3$ & 34.91 / 0.0023 & 30.80 / 0.0033 & 29.43 / 0.0041 & 29.40 / 0.0069 & 34.71 / 0.0046  \\
\hline
p-value & $\times 3$ & 0.0015       & 0.0003         & 8.98e-8       & 2.44e-9        & 4.64e-6           \\
\hline
\hline
NLSN     & $\times 4$ & 32.59 / 0.0027 & 28.87 / 0.0024 & 27.78 / 0.0045 & 26.96 / 0.0060 & 31.27 / 0.0062  \\
HAT     & $\times 4$ & 32.65 / 0.0027 & 28.86 / 0.0024 & 27.76 / 0.0045 & 26.97 / 0.0060 & 31.30 / 0.0062  \\
ACLA    & $\times 4$ & 32.68 / 0.0035 & 28.98 / 0.0033 & 27.86 / 0.0051 & 27.12 / 0.0055 & 31.53 / 0.0073  \\
\hline
p-value & $\times 4$ & 0.0012          & 0.0007          & 0.0005        & 4.57e-12        & 3.62e-9          \\
\hline
\end{tabular}
}
\end{table*}
\begin{table}[!htbp]
\centering
    \caption{PSNR (mean/std) results comparison with p-value between ACLA and SCUNet for image denoising}
    \vspace{-3mm}
        \label{table:DN_q}
        \resizebox{1.0\columnwidth}{!}{

\begin{tabular}{|c|c|c|c|c|}
\hline
Methdos & $\sigma$ & KCLDAk24        & BSD68           & Urban100         \\
\hline
SCUNet   & 10       & 37.41 / 0.0076 & 36.52 / 0.0077 & 36.87 / 0.0082    \\
ACLA    & 10       & 37.52 / 0.0066 & 36.65 / 0.0087 & 36.97 / 0.0085  \\
\hline
p-value & 10       & 3.95e-13        &8.19e-11       & 7.97e-10        \\
\hline
\hline
SCUNet   & 30       & 31.99 / 0.0059 & 30.71 / 0.0061 & 31.91 / 0.0073  \\
ACLA    & 30       & 32.10 / 0.0057 & 30.83 / 0.0073 & 31.99 / 0.0059  \\
\hline
p-value & 30       & 6.75e-9         &7.54e-9       & 2.50e-10         \\
\hline
\hline
SCUNet   & 50       & 29.65 / 0.0053 & 28.35 / 0.0084 & 29.48 / 0.0069  \\
ACLA    & 50       & 29.78 / 0.0070 & 28.47 / 0.0088 & 29.63 / 0.0079  \\
\hline
p-value & 50       & 4.23e-12         &5.29e-12         & 2.97e-11        \\
\hline
\hline
SCUNet   & 70       & 28.21 / 0.0079 & 26.85 / 0.0084 & 27.90 / 0.0062  \\
ACLA    & 70       & 28.33 / 0.0084 & 26.99 / 0.0086 & 27.99 / 0.0086  \\
\hline
p-value & 70       & 3.87e-12        & 9.31e-11          & 1.77e-11          \\
\hline
\end{tabular}

}
\end{table}
\subsection{Image Denoising}
We also evaluate ACLA module on standard benchmarks, KCLDAk24, BSD68~\cite{martin2001database}, and Urban100~\cite{huang2015single}, for image denoising. The noisy images are created by adding AWGN noises with $\sigma = 10, 30, 50, 70$. We compare our approach with four baseline methods, DnCNN~\cite{zhang2017beyond}, MemNet~\cite{tai2017memnet}, RNAN~\cite{zhang2019residual}, PANet~\cite{mei2020pyramid}, SwinIR\cite{liang2021swinir}, SCUNet~\cite{zhang2022practical}, and Restormer~\cite{zamir2022restormer}. Note that the results of SwinIR, SCUNet, and Restormer reported in \cite{zhang2022practical, zamir2022restormer} are obtained by models trained on DIV2K \cite{timofte2017ntire} and Flick2K \cite{EDSR}. In our experiments, we train SwinIR, SCUNet, and Restormer on DIV2K with the same settings as ACLA for fair comparisons. A 16-layer EDSR is used as the baseline CNN backbone, and ACLA modules are inserted into such neural backbone. We use PNSR as the metric to evaluate different methods. As shown in Table~\ref{tab:results_psnr_denoise_rgb}, our methods achieve remarkable improvements on all benchmarks with all noise levels. \textbf{The average improvements of PSNR over the top baseline SCUNet for noise levels 10, 30, 50, and 70 are 0.113 dB, 0.103 dB, 0.133 dB, and 0.110 dB.} To verify that such improvement is statistically significant and out of the range of error, we train ACLA and the current SOTA method SCUNet on different noise levels ten times with different seeds for random initialization of the networks. The mean and standard deviation of different runs are shown in Table~\ref{table:DN_q}. Then we perform t-test between the results of ACLA and SCUNet on all benchmark datasets with all the noise levels. The largest p-value among all the datasets and noise levels is 7.54e-9 $\ll 0.05$ on BSD68 with noise level of 30, suggesting that the improvement of ACLA over SCUNet for image denoising is statistically significant.

\subsection{Image Compression Artifacts Reduction}
For the task of image compression artifacts reduction (CAR), we compare our methods with DnCNN~\cite{zhang2017beyond}, RNAN~\cite{zhang2019residual}, PANet~\cite{mei2020pyramid}, and SwinIR~\cite{liang2021swinir}. All methods are evaluated on LIVE1~\cite{sheikh2005live} and Classic5~\cite{foi2007pointwise}. To obtain the low-quality compressed images, we follow the standard JPEG compression process and use the MATLAB JPEG encoder with quality $q = 10, 20, 30, 40$.  For a fair comparison, the results are only evaluated on the Y channel in the YCbCr Space. We also use PSNR as the metric to evaluate different methods. The results are shown in Table~\ref{table:CAR_results}, where a 16-layer EDSR is used as the baseline CNN backbone. It can be observed that ACLA boosts the performance of the CNN backbone and surpass other baseline methods on all the benchmarks at different JPEG compression qualities. \textbf{The average improvements of PSNR over the top baseline SwinIR for compression quality 10, 20, 30, and 40 are 0.115 dB, 0.100 dB, 0.115 dB, and 0.105 dB}. To verify that such improvement is statistically significant and out of the range of error, we train ACLA and the current SOTA method SwinIR for different compression qualities ten times with different seeds for random initialization of the networks. The mean and standard deviation of different runs are shown in Table~\ref{table:CAR_q}. Then we perform t-test between the results of ACLA and SwinIR on all benchmark datasets with all the compression qualities. The largest p-value of the t-test among all the datasets and compression qualities is $0.0024 \ll 0.05$ on Classic5 with compression quality 30, which is much less than 0.05, suggesting that the improvement of ACLA over SwinIR for image compression artifacts reduction is statistically significant.

\subsection{Image Demosaicing}
For the task of image demosaicing, the evaluation is conducted on Kodak24, McMaster~\cite{zhang2017learning}, BSD68, and Urban100, following the settings in RNAN~\cite{zhang2019residual}. We compare our methods with IRCNN~\cite{zhang2017learning}, RNAN~\cite{zhang2019residual}, and PANet~\cite{mei2020pyramid}. A 16-layer EDSR serves as the baseline CNN model. PSNR is used as the metric to evaluate different methods. As shown in Table \ref{tab:results_psnr_demosaic_rgb}, ACLA always yields the best reconstruction result for image demosaicing. \textbf{The average improvement of PSNR over the top baseline PANet for image demosaicing is 0.135 dB.} To verify that such improvement is statistically significant and out of the range of error, we train ACLA and the current SOTA method PANet ten times with different seeds for random initialization of the networks. The mean and standard deviation of different runs are shown in Table~\ref{table:DM_q}. Then we perform t-test between the results of ACLA and PANet on all benchmark datasets with all the compression qualities. The largest p-value of the t-test among all the datasets is 7.63e-10$\ll 0.05$ on McMaster18, suggesting that the improvement of ACLA over PANet for image demosaicing is statistically significant.

\begin{table*}
\centering
    \caption{PSNR (mean/std) results comparison with p-value between ACLA and SwinIR for image compression artifacts reduction}

    \vspace{-3mm}
        \label{table:CAR_q}
        \resizebox{1.0\textwidth}{!}{
\begin{tabular}{|c|cccc|cccc|}
\hline
\multirow{2}{*}{Method} & \multicolumn{4}{c|}{LIVE1}                                                                                                       & \multicolumn{4}{c|}{Classic5}                                                                                                    \\ \cline{2-9}
                        & \multicolumn{1}{c|}{10}             & \multicolumn{1}{c|}{20}             & \multicolumn{1}{c|}{30}             & 40             & \multicolumn{1}{c|}{10}             & \multicolumn{1}{c|}{20}             & \multicolumn{1}{c|}{30}             & 40             \\ \hline
SwinIR                  & \multicolumn{1}{c|}{29.74 / 0.0067} & \multicolumn{1}{c|}{32.13 / 0.0073} & \multicolumn{1}{c|}{33.57 / 0.0067} & 34.63 / 0.0091 & \multicolumn{1}{c|}{30.06 / 0.0061} & \multicolumn{1}{c|}{32.43 / 0.0075} & \multicolumn{1}{c|}{33.55 / 0.0082} & 34.42 / 0.0083 \\
ACLA                    & \multicolumn{1}{c|}{29.83 / 0.0081} & \multicolumn{1}{c|}{32.25 / 0.0066} & \multicolumn{1}{c|}{33.68 / 0.0073} & 35.55 / 0.0089 & \multicolumn{1}{c|}{30.20 / 0.0077} & \multicolumn{1}{c|}{32.51 / 0.0079} & \multicolumn{1}{c|}{33.67 / 0.0083} & 34.55 / 0.0082 \\ \hline
p-value                 & \multicolumn{1}{c|}{8.79e-12}       & \multicolumn{1}{c|}{0.0003}         & \multicolumn{1}{c|}{0.0002}         & 5.75e-9        & \multicolumn{1}{c|}{8.19e-8}        & \multicolumn{1}{c|}{0.0019}         & \multicolumn{1}{c|}{0.0024}         & 0.0001         \\ \hline
\end{tabular}

}
\end{table*}
\begin{table}[!htbp]
\centering
    \caption{PSNR (mean/std) results comparison with p-value between ACLA and PANet for image demosaicing}
    \vspace{-3mm}
        \label{table:DM_q}
        \resizebox{1.0\columnwidth}{!}{
\begin{tabular}{|c|c|c|c|c|}
\hline
Methods & McMaster18      & Kodak24         & BSD68           & Urban100         \\
\hline
PANet   & 40.00 / 0.0090 & 43.29 / 0.0118 & 42.86 / 0.0095 & 40.50 / 0.0112  \\
ACLA    & 40.13 / 0.0131 & 43.42 / 0.0116 & 43.00 / 0.0117 & 40.63 / 0.0125  \\
\hline
p-value & 7.63e-10        &5.95e-12         &8.41e-11         & 5.97e-10          \\
\hline
\end{tabular}

}
\end{table}
% \subsection{Improvement Significance Analysis}
% To verify that the improvement of our proposed ACLA on existing methods is statistically significant and out of the range of error, we train both ACLA and the current SOTA method ten times with different seeds for random initialization of the networks. Besides, we perform t-test between the results of ACLA and the results of the current SOTA method on each test dataset to see if the improvement of ACLA is statistically significant. The results for single image super-resolution are displayed in Table~\ref{table:SR_q}. The comparison is performed between ACLA and NLSN. We can see that ACLA consistently improves the PSNR results on different test datasets.
% % The standard deviation of PSNR results on all test datasets is less than 0.001.
% The largest p-value is 0.0021 on Set 5, which is less than 0.05. This suggests that the improvement of ACLA over NLSN is statistically significant with $p\ll 0.05$, and it is not caused by random error. For image denoising, image compression artifacts reduction, and image demosaicing, we compare ACLA with the current SOTA, PANet. The results are shown in Table~\ref{table:DN_q}, Table~\ref{table:CAR_q}, and Table~\ref{table:DM_q} respectively. We can see that ACLA consistently improves the performance on different test datasets. The p-value for all these three tasks is less than 0.05, suggesting the statistically significant improvement of ACLA over PANet.

\subsection{Ablation Study and Discussion}
\vspace{-1mm}
\label{sec:ablation_study}
\subsubsection{ACLA vs. Non-Local Attention}
To verify the effectiveness of our proposed methods, we compare ACLA with the vanilla Non-Local (NL) attention \cite{wang2018non} defined in Equation~(\ref{eq:original_nl}) and vanilla Cross-Layer Non-Local (CLNL) attention defined in Equation~(\ref{eq:cl_nl}) in terms of computational efficiency and performance. The CLNL follows the formulation in equation~(\ref{eq:cl_nl}). The comparison is performed on Set 5 and Set 14 for $2\times$ single image super-resolution with EDSR backbone. The NL and CLNL modules are inserted evenly after every 8th residual block. All the FLOPs in our ablation study are calculated for an input size of $48\times48$. Results are presented in Table~\ref{table:efficiency}. It can be observed that, with less computation cost, ACLA achieve much better performance compared to standard NL and CLNL modules.

\begin{table}[!htbp]
    \centering
    \caption{Efficiency comparison with Non-Local attention on Set5}
    \vspace{-3mm}
        \label{table:efficiency}
        %\vspace{-0.7em}
        \resizebox{0.4\textwidth}{!}{
\begin{tabular}{|c|c|c|c|c|}
\hline
Method        & FLOPs(G) & Params(M)& Set 5 & Set 14 \\ \hline
EDSR          & 93.97    & 40.73    & 38.11 & 33.92 \\
NL            & 109.38   & 43.56    & 38.15 & 34.00 \\
CLNL          & 122.67   & 45.87    & 38.14 & 34.05 \\
ACLA (Ours)   & 96.97    & 42.29    & 38.39 & 34.24 \\ \hline
\end{tabular}
        }
\end{table}
\subsubsection{ACLA vs. State-of-the-art Attention Modules}
\label{sec:ablation-sota-attention}
\vspace{-1mm}
In this subsection, we compare ACLA with several state-of-the-art attention modules that are widely used in the CV community, including Squeeze-and-Excitation (SE) \cite{hu2018squeeze} attention and Multi-Head Attention (MHA) \cite{bello2019attention}. SE models interdependencies between the channels of the convolutional features by re-weighting the channel-wise responses using soft self-attention. MHA is in fact a variant of self-attention from the NLP domain. Specifically, MHA can be regarded as a special non-local attention module that takes account of the relative position information. We insert four SE blocks and four MHA blocks evenly to the EDSR backbone, forming the baseline methods EDSR + SE and EDSR + MHA respectively in Table~\ref{table:other_attention}. The comparison is performed for $2\times$ single-image super-resolution on Set 5 and Set 14. The comparative results are shown in Table~\ref{table:other_attention}. Although MHA and SE bring improvements over the EDSR baseline, the best results are achieved by our proposed ACLA. Furthermore, we achieve even better performance by inserting a SE block after each ACLA module, as shown in the last row of Table~\ref{table:other_attention}.

\begin{table}[htb]
    \centering
    \caption{Efficiency and performance comparison with Squeeze-and-Excitation (SE) attention and Multi-Head Attention (MHA)}
    \vspace{-3mm}

        \label{table:other_attention}
        %\vspace{-0.7em}
        \resizebox{0.5\textwidth}{!}{
\begin{tabular}{|c|c|c|c|c|}
\hline
Methods           & FLOPs(G) & Params(M) & Set 5 & Set 14 \\ \hline
EDSR              & 93.97    & 40.73     & 38.11 & 33.92  \\
EDSR + MHA        & 100.21   & 42.17     & 38.23 & 34.01  \\
EDSR + SE         & 96.14    & 41.79     & 38.19 & 34.03  \\
EDSR + ACLA       & 96.97    & 42.29     & 38.39 & 34.24  \\
EDSR + ACLA + SE  & 99.32    & 43.47     & 38.40 & 34.27  \\ \hline
\end{tabular}
        }
\end{table}

\subsubsection{Ablation Study on the Two Adaptive Designs of ACLA}
\label{sec:ablation-two-adaptive-designs}
In Section~\ref{sec:ACLA}, two adaptive designs are proposed and applied to our ACLA module. The first adaptive design is to select an adaptive number of keys at each layer for non-local attention, and the second adaptive design is to search for optimal insert positions of ACLA modules. To verify the effectiveness of these two adaptive designs in ACLA, we design a baseline method termed Cross-Layer Attention (CLA). Different from ACLA, the insert positions for CLA are fixed. In our experiment, we insert four CLA modules evenly after every 8th residual block in the EDSR backbone. Each query of CLA refers to a fixed number, that is $K$, of keys from each previous layer. Thus, the formulation of CLA is $y_i^{j} = \frac{1}{\mathcal{C}(x^j)} \sum_{l=1}^{j} \sum_{k=1}^{K}  f(x_i^j)g(x^l(p_i+\Delta p_{ik}))$. Compared to the formulation of ACLA in Equation~(\ref{eq:masked_csnl_search}), CLA takes the architecture parameters $s_l$ and $m_{i,k}^{j,l}$ as $1$.

To separately verify the effectiveness of the two adaptive designs in ACLA. We further design two baseline modules based on CLA, that are CLA-I and CLA-K. CLA-I stands for CLA with the search for insert positions as that in ACLA. CLA-K stands for CLA which selects an adaptive number of keys at each layer as that in ACLA.

We perform comparison between ACLA, CLA-I, CLA-K, and CLA on Set 5 and Set 14 for $\times2$ single image super-resolution with EDSR backbone. The comparative results are shown in Table~\ref{table:ablation_NAS}. It can be observed that each adaptive design brings improvement on the baseline CLA. ACLA, as a combination of the two adaptive designs, renders better performance than each individual adaptive design.
\begin{table}[htb]
    \centering
    \caption{Ablation study on the effectiveness of insertion position search and adaptive key selection}
\vspace{-2mm}
        \label{table:ablation_NAS}
        %\vspace{-0.7em}
        \resizebox{0.9\columnwidth}{!}{
\begin{tabular}{|c|c|c|c|c|}
\hline
Method  & FLOPs(G) & Params(M)& Set 5  & Set 14  \\ \hline
CLA     & 96.93    & 42.13    & 38.27 & 34.07 \\
CLA-I   & 96.93    & 42.13    & 38.33 & 34.13 \\
CLA-K   & 96.87    & 42.29    & 38.32 & 34.15 \\
ACLA    & 96.97    & 42.29    & 38.39 & 34.24 \\ \hline
\end{tabular}
        }
\end{table}
\begin{table}[htbp]

\centering
\caption{Ablation study on number of sampled keys in ACLA on Set5}
\vspace{-2mm}
\label{table:K_CLA}
\resizebox{0.9\columnwidth}{!}{
\begin{tabular}{|c|c|c|c|c|c|}
\hline
Method & $K$  & FLOPs(G) & Params(M)& Set 5 & Set 14 \\ \hline
ACLA   & 8    & 96.78    & 42.18    & 38.35 & 34.16\\
ACLA   & 16   & 96.97    & 42.29    & 38.39 & 34.24\\
ACLA   & 32   & 97.56    & 42.41    & 38.38 & 34.25\\
ACLA   & 64   & 98.03    & 42.69    & 38.39 & 34.23\\
ACLA   & 128  & 99.17    & 43.02    & 38.37 & 34.22\\
ACLA   & 256  & 100.59   & 43.97    & 38.37 & 34.24\\ \hline
\end{tabular}
}
\end{table}
\subsubsection{Ablation Study on the Number of Selected keys $K$ in ACLA}
% As discussed in Section~\ref{sec::CLA}, CLA selects a fixed number of keys, that is $K$, at each layer for non-local attention. To verify that a small $K$ is sufficient for competitive performance, we perform experiments on CLA with different values of $K$.

To verify that a small $K$, which is the maximal number of sampled keys, is sufficient for competitive performance, we compare the performance of ACLA with different values of $K$. The comparison is performed on Set 5 and Set 14 for $\times2$ single image super-resolution with EDSR backbone. The results are shown in Table~\ref{table:K_CLA}. With increased $K$, the performance of ACLA does not constantly improve. ACLA with $K=16$ can already achieve comparable performance to those with larger $K$. This is also consistent with previous studies \cite{zhang2015survey, elad2006image} on the power of sparse representation learning for image restoration.

\begin{figure*}[h]
\begin{subfigure}{1.0\textwidth}
  \centering
  \includegraphics[width=0.925\linewidth]{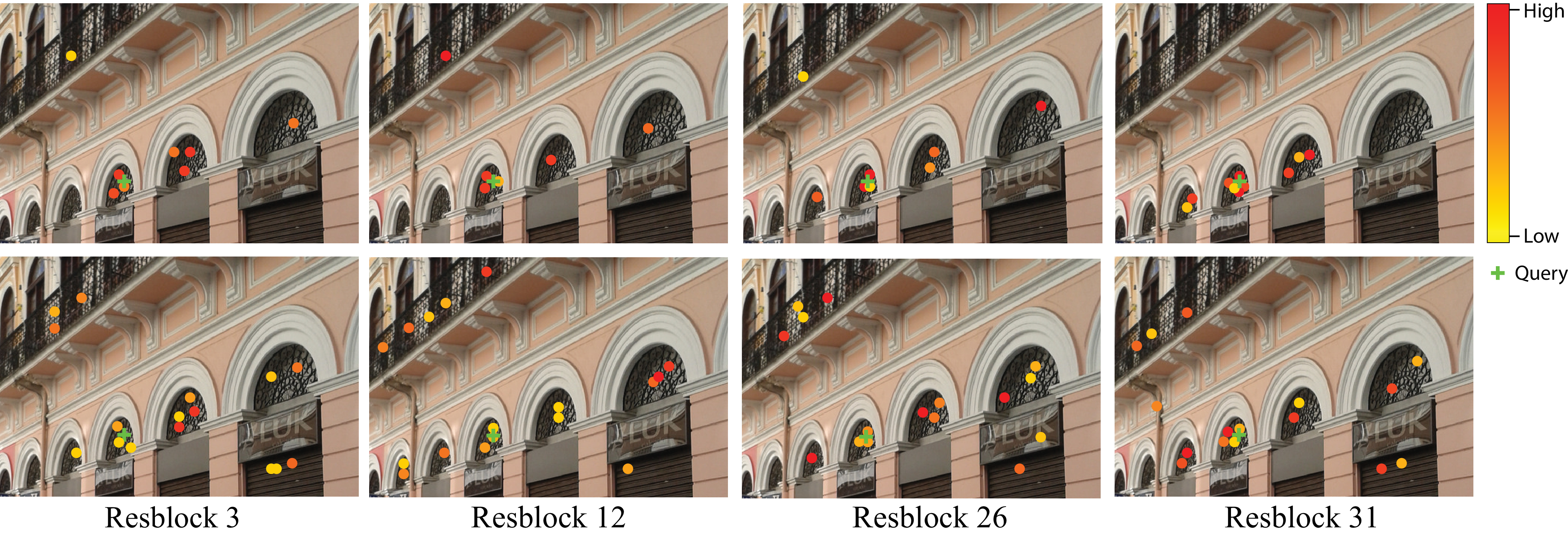}
  \caption{}
  \label{fig:sfig1}
\end{subfigure}%

\begin{subfigure}{1.0\textwidth}
  \centering
  \includegraphics[width=0.925\linewidth]{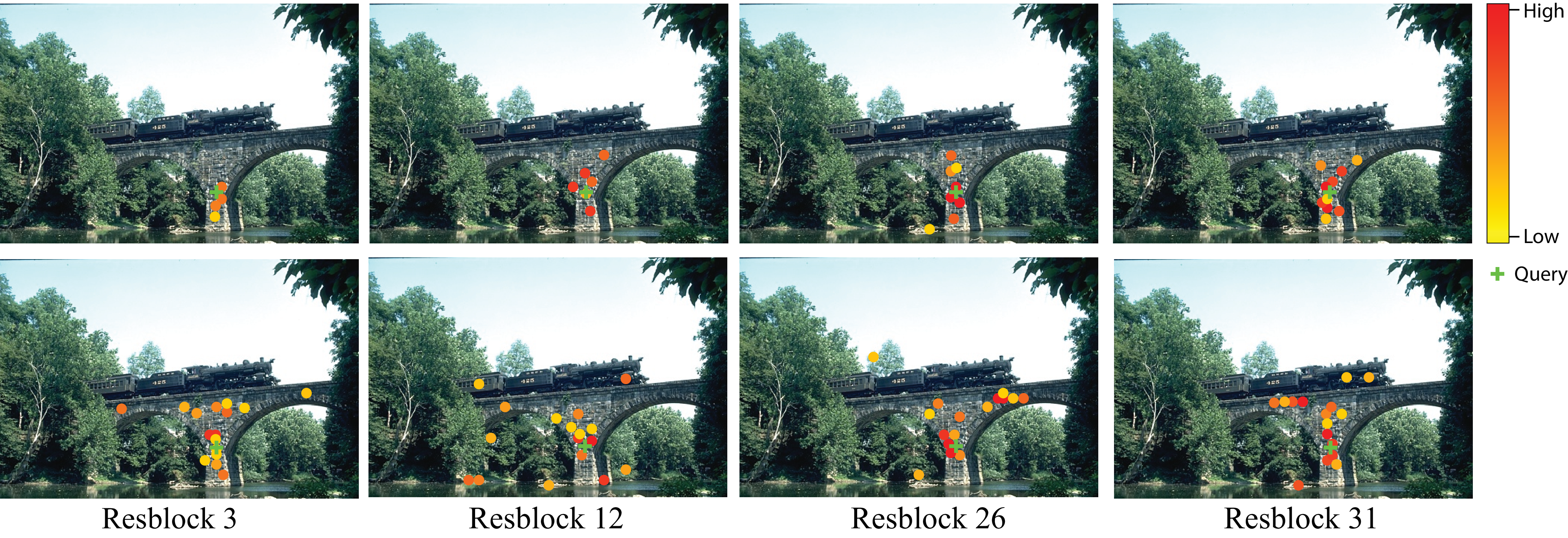}
  \caption{}
  \label{fig:sfig2}
\end{subfigure}

\begin{subfigure}{1.0\textwidth}
  \centering
  \includegraphics[width=0.925\linewidth]{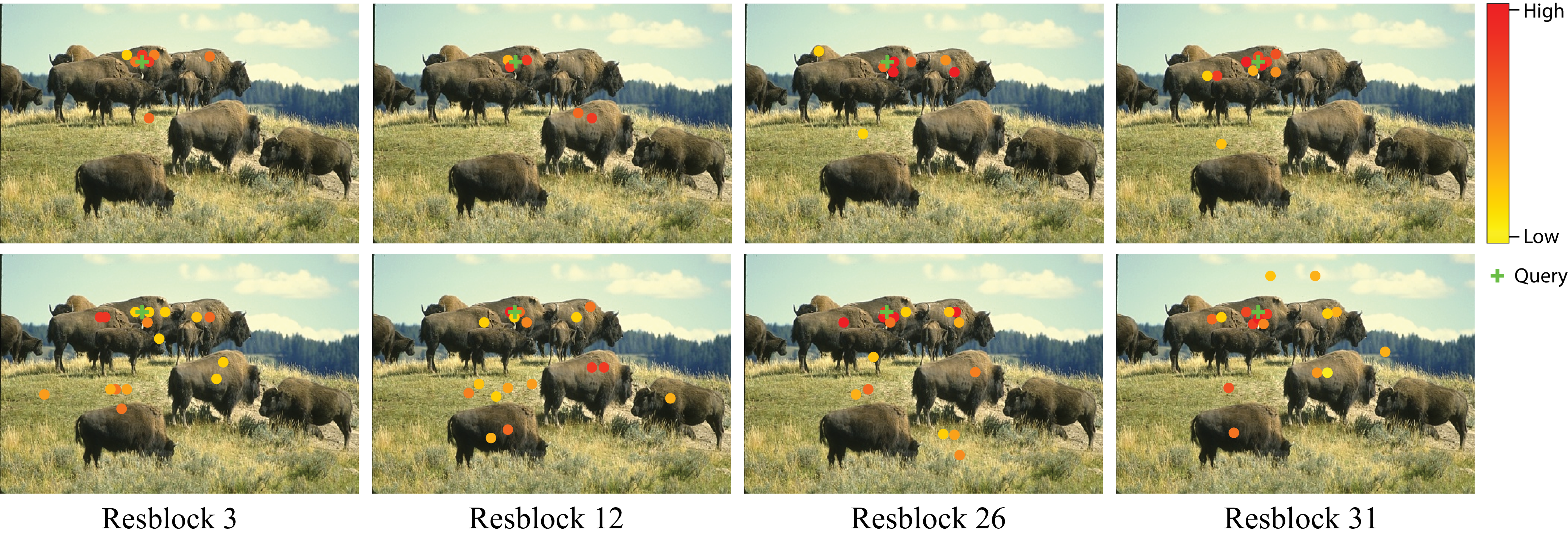}
  \caption{}
  \label{fig:sfig3}
\end{subfigure}
\caption{Visualization of selected keys by ACLA  }
\label{fig:more_visualization}
\end{figure*}

\subsubsection{Ablation Study on Automatic Search for Referred Layers in ACLA}
\label{sec:ablation-referred-layers}
To verify the effectiveness of the automatic search for referred layers in ACLA, we compare ACLA against baselines where queries in each inserted ACLA module refer to keys from the output of a fixed number of preceding layers, which are termed Fixed-Layer ACLA. The comparison is performed on Set 5 and Set 14 for $\times2$ single image super-resolution using EDSR backbone with 32 residual blocks. The experiment settings are the same as reported in Section \ref{sec:settings} of our paper.  The ACLA modules in Fixed-Layer ACLA are inserted to the same positions in the EDSR backbone as ACLA. However, queries in each ACLA module only refer to keys from the outputs of the  previous $i$ residual blocks. When $i = 1$, ACLA with $1$ referred layer only refers to keys from the output of the same residual block. When the number of residual blocks previous to an ACLA module is less than $i$, queries in that ACLA module refer to keys from the outputs of all previous residual blocks. The results in Table~\ref{table:ACLA_refer} show that ACLA outperforms Fixed-Layer ACLA with different number of referred layers. In addition, ACLA enjoys less FLOPs and parameter number than the top baseline in Table~\ref{table:ACLA_refer}, evidending the effectiveness and efficiency of automatic search for referred layers in ACLA.

\begin{table}[htb]
    \centering
    \caption{Ablation study on Automatic Search for Referred Layers in ACLA}
    \vspace{-3mm}
        \label{table:ACLA_refer}
        %\vspace{-0.7em}
        \resizebox{1\columnwidth}{!}{
\begin{tabular}{|c|c|c|c|c|c|}
\hline
Methods & $\#$ Referred Layers (i) & FLOPs(G) & Params(M) & Set 5 & Set 14 \\ \hline
EDSR    & -      & 93.97    & 40.73     & 38.11 & 33.92  \\
Fixed-Layer ACLA    & 1      & 94.52    & 40.92     & 38.26 & 34.13  \\
Fixed-Layer ACLA    & 2      & 95.13    & 41.23     & 38.31 & 34.15  \\
Fixed-Layer ACLA    & 4      & 96.22    & 41.89     & 38.33 & 34.17  \\
Fixed-Layer ACLA    & 8      & 101.98   & 43.94     & \underline{38.34} & \underline{34.19}  \\
Fixed-Layer ACLA    & 16     & 109.79   & 47.12     & 38.32 & 34.17  \\
Fixed-Layer ACLA    & 32     & 125.67   & 54.89     & \underline{38.34} & 34.18  \\
ACLA    & -      & 96.97    & 42.29     & \textbf{38.39} & \textbf{34.24}  \\
\hline
\end{tabular}
        }
\end{table}
% \begin{table}[htb]
%     \centering
%     \caption{Ablation study on the number of referred layers in CLA}
%     \vspace{-3mm}
%       \label{table:CLA_refer}
%       %\vspace{-0.7em}
%       \resizebox{1\columnwidth}{!}{
% \begin{tabular}{|c|c|c|c|c|c|c|}
% \hline
% Methods  & Layers & Set 5 & Set 14 & B100  & Urban100 & Manga109  \\
% \hline
% EDSR     & -               & 38.11 & 33.92  & 32.32 & 32.93    & 39.10     \\

% CLA & 1               & 38.16 & 34.00  & 32.35 & 33.03    & 39.17     \\

% CLA & 2               & 38.18 & 34.04  & 32.37 & 33.11    & 39.20     \\

% CLA & 3               & 38.23 & 34.06  & 32.41 & 33.26    & 39.23     \\

% CLA & 4               & 38.24 & 34.08  & 32.41 & 33.28    & 39.23     \\
% \hline
% \end{tabular}
%       }
% \end{table}

\subsubsection{Inference Time Comparison}

We compare the inference time between our proposed ACLA and previous state-of-the-art methods based on attention modules. The running time is the average of 1000 runs on the input of size $48\times48$. We evaluate the running time on a single Tesla V100 16G. We compare our proposed methods with HAN\cite{niu2020single}, SAN \cite{dai2019second}, and NLSN \cite{mei2021image}, which are also attention-based methods for single image super-resolution. As shown in Table~\ref{table:inf_time}, EDSR+ACLA achieves  better performance than competing methods with less inference time.
\begin{table}[!htb]
    \centering
    \caption{Inference time comparison}
    \vspace{-2mm}
        \label{table:inf_time}

        \resizebox{0.9\columnwidth}{!}{
\begin{tabular}{|c|c|c|c|c|}
\hline
          & HAN   & SAN & NLSN & EDSR+ACLA    \\
\hline
Set 5 (PSNR)            & 38.27 & 38.31 & 38.34     & \textbf{38.39}  \\
Time(ms)         & 38.9  & 61.2 & 20.8      & \textbf{19.8}  \\
\hline
\end{tabular}
        }
\end{table}

\subsubsection{Analysis on Search Results}
\vspace{-1mm}
We summarize the value of $\lambda$, i.e., hyper-parameter that controls the magnitude of the inference cost term, for different tasks in Table~\ref{table:lamda_setting}. The insert positions of ACLA in the searched models are also shown in the same table. For experiments with EDSR \citep{lim2017enhanced}, ACLA modules are inserted after each residual block in the super network. Note that EDSR with 16 residual blocks is used for image denoising, image demosaicing, and image compression artifacts reduction.
\begin{table}[!htbp]
    \centering
    \caption{Search settings for ACLA in different image restoration tasks}
    \vspace{-3mm}
        \label{table:lamda_setting}
        %\vspace{-0.7em}
        \resizebox{1\columnwidth}{!}{
\begin{tabular}{|c|c|c|c|}
\hline
Task & Backbone  & Value of $\lambda$    & Insert Positions (block number) \\ \hline
Super-Resolution & 32-block EDSR    & 0.15 & 3, 12, 26, 31, 32 \\
Denoising      &16-block EDSR&  0.25   & 2, 5, 9, 12, 15    \\
Demosaicing    &16-block EDSR&  0.3  & 2, 5, 11, 13, 16     \\
Artifacts  Reduction   &16-block EDSR&  0.3  & 2, 7, 9, 13, 16  \\
\hline
\end{tabular}}
\end{table}

\subsection{Visualization of Selected Keys}
\label{subsec:visualization}
We present more examples of visualization of selected keys by ACLA in Figure~\ref{fig:more_visualization} to demonstrate the superiority of our method in searching for informative keys for the query feature. The visualization is based on our results for $2\times$ image super-resolution. Similar to Figure~\ref{fig:key_sample_vis}, the first row shows the positions of the keys selected by ACLA with $K=16$. For comparison, the positions of keys with top-$16$ attention weights following the vanilla CLNL attention formulation in Equation~(\ref{eq:cl_nl}) are displayed in the second row. From left to right are the sampled key positions from the 3rd, 12th, 26th, and 31st residual blocks.

The visualization results show that ACLA adaptively selects semantically similar keys for the query feature, and its vanilla counterpart CLNL lacks such capability. For instance, in Figure~\ref{fig:key_sample_vis}, the query is from the ear of the elephant on the right side. With ACLA, $60\%$ of the selected keys across are also from the ear of the same elephant. Besides, among the keys selected outside the ear of the same elephant, 5 out of 11 are from the ear of the elephant on the left, which has similar textures as the ear of the elephant on the right. While with CLNL, only $39\%$ of the selected keys are from the ear of the elephant on the right. Similar observations can also be found in Figure~\ref{fig:more_visualization}. In Figure~\ref{fig:more_visualization}~(a), we pick a query point from the frame structure at the top of a gate. With ACLA, $90\%$ of the keys selected are distributed on the frame structures at the top of the gates. While with CLNL, positions from the gates and the frame structure at the balcony are also given high attention weights. Only $61\%$ of the selected keys are distributed on the frame structures at the top of gates, which may limit the power of attention modules. Similar observations can also be found Figure~\ref{fig:more_visualization}~(b) and Figure~\ref{fig:more_visualization}~(c). In Figure~\ref{fig:more_visualization}~(b), the query is from the bridge in the middle of the image. All the keys selected by ACLA are also from the bridge. In Figure~\ref{fig:more_visualization}~(c), the query is from the back of a yak. Most of the keys selected by ACLA are also located on the body of yaks. While as shown in the second row, CLNL even assigns large attention weights to positions from the grass and the background. Such observations strongly demonstrate the power of ACLA in searching for informative keys across different layers.

\section{Conclusions}
\label{sec:conclusion}
In this paper, we propose Adaptive Cross-Layer Attention, or ACLA, which searches for informative keys across different layers for each query feature in attention modules for image restoration. ACLA features two adaptive designs, selecting an adaptive number of keys at each layer and searching for insert positions of the ACLA modules. In particular, each query feature selects adaptive keys at their referred layers. A neural architecture search method is used to search for the insert positions of the ACLA modules so that the neural network with ACLA modules is compact with competitive performance, which also enables automatic search for referred layers for each query feature. Experiments on image restoration tasks including single-image super-resolution, image denoising, image compression artifacts reduction, and image demosaicing validate the effectiveness and efficiency of the proposed ACLA module.

% or


% \appendices

% % you can choose not to have a title for an appendix
% % if you want by leaving the argument blank
% \section{}

% \ifCLASSOPTIONcompsoc
%   % The Computer Society usually uses the plural form
%   \section*{Acknowledgments}
% \else
%   % regular IEEE prefers the singular form
% \fi


\ifCLASSOPTIONcaptionsoff
  \newpage
\fi



\bibliographystyle{IEEEtran}
\bibliography{main}

\begin{thebibliography}{10}
\providecommand{\url}[1]{#1}
\csname url@samestyle\endcsname
\providecommand{\newblock}{\relax}
\providecommand{\bibinfo}[2]{#2}
\providecommand{\BIBentrySTDinterwordspacing}{\spaceskip=0pt\relax}
\providecommand{\BIBentryALTinterwordstretchfactor}{4}
\providecommand{\BIBentryALTinterwordspacing}{\spaceskip=\fontdimen2\font plus
\BIBentryALTinterwordstretchfactor\fontdimen3\font minus
  \fontdimen4\font\relax}
\providecommand{\BIBforeignlanguage}[2]{{%
\expandafter\ifx\csname l@#1\endcsname\relax
\typeout{** WARNING: IEEEtran.bst: No hyphenation pattern has been}%
\typeout{** loaded for the language `#1'. Using the pattern for}%
\typeout{** the default language instead.}%
\else
\language=\csname l@#1\endcsname
\fi
#2}}
\providecommand{\BIBdecl}{\relax}
\BIBdecl

\bibitem{zhang2019residual}
Y.~Zhang, K.~Li, K.~Li, B.~Zhong, and Y.~Fu, ``Residual non-local attention
  networks for image restoration,'' in \emph{International Conference on
  Learning Representations}, 2019.

\bibitem{liu2018non}
D.~Liu, B.~Wen, Y.~Fan, C.~C. Loy, and T.~S. Huang, ``Non-local recurrent
  network for image restoration,'' in \emph{Advances in Neural Information
  Processing Systems}, 2018, pp. 1673--1682.

\bibitem{zhang2017learning}
K.~Zhang, W.~Zuo, S.~Gu, and L.~Zhang, ``Learning deep cnn denoiser prior for
  image restoration,'' in \emph{CVPR}, 2017.

\bibitem{fan2019scale}
Y.~Fan, J.~Yu, D.~Liu, and T.~S. Huang, ``Scale-wise convolution for image
  restoration,'' \emph{arXiv preprint arXiv:1912.09028}, 2019.

\bibitem{lai2017deep}
W.-S. Lai, J.-B. Huang, N.~Ahuja, and M.-H. Yang, ``Deep laplacian pyramid
  networks for fast and accurate super-resolution,'' in \emph{CVPR}, 2017.

\bibitem{tai2017memnet}
Y.~Tai, J.~Yang, X.~Liu, and C.~Xu, ``Memnet: A persistent memory network for
  image restoration,'' in \emph{Proceedings of the IEEE international
  conference on computer vision}, 2017, pp. 4539--4547.

\bibitem{zhang2017beyond}
K.~Zhang, W.~Zuo, Y.~Chen, D.~Meng, and L.~Zhang, ``Beyond a gaussian denoiser:
  Residual learning of deep cnn for image denoising,'' \emph{TIP}, 2017.

\bibitem{buades2005non}
A.~Buades, B.~Coll, and J.-M. Morel, ``A non-local algorithm for image
  denoising,'' in \emph{CVPR}, 2005.

\bibitem{zoran2011learning}
D.~Zoran and Y.~Weiss, ``From learning models of natural image patches to whole
  image restoration,'' in \emph{2011 International Conference on Computer
  Vision}.\hskip 1em plus 0.5em minus 0.4em\relax IEEE, 2011, pp. 479--486.

\bibitem{zontak2013separating}
M.~Zontak, I.~Mosseri, and M.~Irani, ``Separating signal from noise using patch
  recurrence across scales,'' in \emph{Proceedings of the IEEE Conference on
  Computer Vision and Pattern Recognition}, 2013, pp. 1195--1202.

\bibitem{wang2018non}
X.~Wang, R.~Girshick, A.~Gupta, and K.~He, ``Non-local neural networks,'' in
  \emph{Proceedings of the IEEE conference on computer vision and pattern
  recognition}, 2018, pp. 7794--7803.

\bibitem{niu2020single}
B.~Niu, W.~Wen, W.~Ren, X.~Zhang, L.~Yang, S.~Wang, K.~Zhang, X.~Cao, and
  H.~Shen, ``Single image super-resolution via a holistic attention network,''
  in \emph{European Conference on Computer Vision}.\hskip 1em plus 0.5em minus
  0.4em\relax Springer, 2020, pp. 191--207.

\bibitem{tay2021omninet}
Y.~Tay, M.~Dehghani, V.~Aribandi, J.~Gupta, P.~Pham, Z.~Qin, D.~Bahri, D.-C.
  Juan, and D.~Metzler, ``Omninet: Omnidirectional representations from
  transformers,'' \emph{arXiv preprint arXiv:2103.01075}, 2021.

\bibitem{EDSR}
B.~Lim, S.~Son, H.~Kim, S.~Nah, and K.~M. Lee, ``Enhanced deep residual
  networks for single image super-resolution,'' in \emph{The IEEE Conference on
  Computer Vision and Pattern Recognition (CVPR) Workshops}, July 2017.

\bibitem{dong2015compression}
C.~Dong, Y.~Deng, C.~Change~Loy, and X.~Tang, ``Compression artifacts reduction
  by a deep convolutional network,'' in \emph{ICCV}, 2015.

\bibitem{zhang2018residual}
Y.~Zhang, Y.~Tian, Y.~Kong, B.~Zhong, and Y.~Fu, ``Residual dense network for
  image super-resolution,'' in \emph{Proceedings of the IEEE conference on
  computer vision and pattern recognition}, 2018, pp. 2472--2481.

\bibitem{haris2018deep}
M.~Haris, G.~Shakhnarovich, and N.~Ukita, ``Deep back-projection networks for
  super-resolution,'' in \emph{Proceedings of the IEEE conference on computer
  vision and pattern recognition}, 2018, pp. 1664--1673.

\bibitem{ahn2018fast}
N.~Ahn, B.~Kang, and K.-A. Sohn, ``Fast, accurate, and lightweight
  super-resolution with cascading residual network,'' in \emph{Proceedings of
  the European Conference on Computer Vision (ECCV)}, 2018, pp. 252--268.

\bibitem{dai2019second}
T.~Dai, J.~Cai, Y.~Zhang, S.-T. Xia, and L.~Zhang, ``Second-order attention
  network for single image super-resolution,'' in \emph{Proceedings of the IEEE
  conference on computer vision and pattern recognition}, 2019, pp.
  11\,065--11\,074.

\bibitem{mei2020pyramid}
Y.~Mei, Y.~Fan, Y.~Zhang, J.~Yu, Y.~Zhou, D.~Liu, Y.~Fu, T.~S. Huang, and
  H.~Shi, ``Pyramid attention networks for image restoration,'' \emph{arXiv
  preprint arXiv:2004.13824}, 2020.

\bibitem{xu2015show}
K.~Xu, J.~Ba, R.~Kiros, K.~Cho, A.~Courville, R.~Salakhudinov, R.~Zemel, and
  Y.~Bengio, ``Show, attend and tell: Neural image caption generation with
  visual attention,'' in \emph{International conference on machine
  learning}.\hskip 1em plus 0.5em minus 0.4em\relax PMLR, 2015, pp. 2048--2057.

\bibitem{chen2017sca}
L.~Chen, H.~Zhang, J.~Xiao, L.~Nie, J.~Shao, W.~Liu, and T.-S. Chua, ``Sca-cnn:
  Spatial and channel-wise attention in convolutional networks for image
  captioning,'' in \emph{Proceedings of the IEEE conference on computer vision
  and pattern recognition}, 2017, pp. 5659--5667.

\bibitem{hu2018squeeze}
J.~Hu, L.~Shen, and G.~Sun, ``Squeeze-and-excitation networks,'' in
  \emph{Proceedings of the IEEE conference on computer vision and pattern
  recognition}, 2018, pp. 7132--7141.

\bibitem{wang2017residual}
F.~Wang, M.~Jiang, C.~Qian, S.~Yang, C.~Li, H.~Zhang, X.~Wang, and X.~Tang,
  ``Residual attention network for image classification,'' in \emph{Proceedings
  of the IEEE conference on computer vision and pattern recognition}, 2017, pp.
  3156--3164.

\bibitem{zhang2018image}
Y.~Zhang, K.~Li, K.~Li, L.~Wang, B.~Zhong, and Y.~Fu, ``Image super-resolution
  using very deep residual channel attention networks,'' in \emph{Proceedings
  of the European Conference on Computer Vision (ECCV)}, 2018, pp. 286--301.

\bibitem{IPT}
H.~Chen, Y.~Wang, T.~Guo, C.~Xu, Y.~Deng, Z.~Liu, S.~Ma, C.~Xu, C.~Xu, and
  W.~Gao, ``Pre-trained image processing transformer,'' in \emph{Proceedings of
  the IEEE/CVF Conference on Computer Vision and Pattern Recognition}, 2021,
  pp. 12\,299--12\,310.

\bibitem{liang2021swinir}
J.~Liang, J.~Cao, G.~Sun, K.~Zhang, L.~Van~Gool, and R.~Timofte, ``Swinir:
  Image restoration using swin transformer,'' \emph{arXiv preprint
  arXiv:2108.10257}, 2021.

\bibitem{zoph2016neural}
B.~Zoph and Q.~V. Le, ``Neural architecture search with reinforcement
  learning,'' in \emph{International Conference on Learning Representations
  (ICLR)}, 2016.

\bibitem{xie2017genetic}
L.~Xie and A.~Yuille, ``Genetic cnn,'' in \emph{Proceedings of the IEEE/CVF
  International Conference on Computer Vision (ICCV)}, 2017.

\bibitem{pham2018efficient}
H.~Pham, M.~Guan, B.~Zoph, Q.~Le, and J.~Dean, ``Efficient neural architecture
  search via parameters sharing,'' in \emph{International Conference on Machine
  Learning (ICML)}, 2018.

\bibitem{liu2018progressive}
C.~Liu, B.~Zoph, M.~Neumann, J.~Shlens, W.~Hua, L.-J. Li, L.~Fei-Fei,
  A.~Yuille, J.~Huang, and K.~Murphy, ``Progressive neural architecture
  search,'' in \emph{Proceedings of the European conference on computer vision
  (ECCV)}, 2018, pp. 19--34.

\bibitem{liu2018darts}
H.~Liu, K.~Simonyan, and Y.~Yang, ``Darts: Differentiable architecture
  search,'' in \emph{International Conference on Learning Representations
  (ICLR)}, 2018.

\bibitem{xie2018snas}
S.~Xie, H.~Zheng, C.~Liu, and L.~Lin, ``Snas: stochastic neural architecture
  search,'' in \emph{International Conference on Learning Representations
  (ICLR)}, 2018.

\bibitem{liu2019auto}
C.~Liu, L.-C. Chen, F.~Schroff, H.~Adam, W.~Hua, A.~L. Yuille, and L.~Fei-Fei,
  ``Auto-deeplab: Hierarchical neural architecture search for semantic image
  segmentation,'' in \emph{Proceedings of the IEEE/CVF Conference on Computer
  Vision and Pattern Recognition (CVPR)}, 2019.

\bibitem{zhang2021dcnas}
X.~Zhang, H.~Xu, H.~Mo, J.~Tan, C.~Yang, L.~Wang, and W.~Ren, ``Dcnas: Densely
  connected neural architecture search for semantic image segmentation,'' in
  \emph{Proceedings of the IEEE/CVF Conference on Computer Vision and Pattern
  Recognition}, 2021, pp. 13\,956--13\,967.

\bibitem{guo2020hierarchical}
Y.~Guo, Y.~Luo, Z.~He, J.~Huang, and J.~Chen, ``Hierarchical neural
  architecture search for single image super-resolution,'' \emph{IEEE Signal
  Processing Letters}, vol.~27, pp. 1255--1259, 2020.

\bibitem{dai2017deformable}
J.~Dai, H.~Qi, Y.~Xiong, Y.~Li, G.~Zhang, H.~Hu, and Y.~Wei, ``Deformable
  convolutional networks,'' in \emph{Proceedings of the IEEE international
  conference on computer vision}, 2017, pp. 764--773.

\bibitem{verelst2020dynamic}
T.~Verelst and T.~Tuytelaars, ``Dynamic convolutions: Exploiting spatial
  sparsity for faster inference,'' in \emph{Proceedings of the IEEE/CVF
  Conference on Computer Vision and Pattern Recognition}, 2020, pp. 2320--2329.

\bibitem{Bengio2013EstimatingComputation}
\BIBentryALTinterwordspacing
Y.~Bengio, N.~L{\'{e}}onard, and A.~Courville, ``{Estimating or Propagating
  Gradients Through Stochastic Neurons for Conditional Computation},'' 8 2013.
  [Online]. Available: \url{http://arxiv.org/abs/1308.3432}
\BIBentrySTDinterwordspacing

\bibitem{fang2020densely}
J.~Fang, Y.~Sun, Q.~Zhang, Y.~Li, W.~Liu, and X.~Wang, ``Densely connected
  search space for more flexible neural architecture search,'' in
  \emph{Proceedings of the IEEE/CVF Conference on Computer Vision and Pattern
  Recognition}, 2020, pp. 10\,628--10\,637.

\bibitem{zhu2020deformable}
X.~Zhu, W.~Su, L.~Lu, B.~Li, X.~Wang, and J.~Dai, ``Deformable {DETR:}
  deformable transformers for end-to-end object detection,'' in \emph{9th
  International Conference on Learning Representations, {ICLR} 2021, Virtual
  Event, Austria, May 3-7, 2021}.\hskip 1em plus 0.5em minus 0.4em\relax
  OpenReview.net, 2021.

\bibitem{timofte2017ntire}
R.~Timofte, E.~Agustsson, L.~Van~Gool, M.-H. Yang, and L.~Zhang, ``Ntire 2017
  challenge on single image super-resolution: Methods and results,'' in
  \emph{Proceedings of the IEEE conference on computer vision and pattern
  recognition workshops}, 2017, pp. 114--125.

\bibitem{lim2017enhanced}
B.~Lim, S.~Son, H.~Kim, S.~Nah, and K.~Mu~Lee, ``Enhanced deep residual
  networks for single image super-resolution,'' in \emph{Proceedings of the
  IEEE conference on computer vision and pattern recognition workshops}, 2017,
  pp. 136--144.

\bibitem{mei2021image}
Y.~Mei, Y.~Fan, and Y.~Zhou, ``Image super-resolution with non-local sparse
  attention,'' in \emph{Proceedings of the IEEE/CVF Conference on Computer
  Vision and Pattern Recognition}, 2021, pp. 3517--3526.

\bibitem{bevilacqua2012low}
M.~Bevilacqua, A.~Roumy, C.~Guillemot, and M.~L. Alberi-Morel, ``Low-complexity
  single-image super-resolution based on nonnegative neighbor embedding,''
  2012.

\bibitem{zeyde2010single}
R.~Zeyde, M.~Elad, and M.~Protter, ``On single image scale-up using
  sparse-representations,'' in \emph{International conference on curves and
  surfaces}, 2010.

\bibitem{martin2001database}
D.~Martin, C.~Fowlkes, D.~Tal, and J.~Malik, ``A database of human segmented
  natural images and its application to evaluating segmentation algorithms and
  measuring ecological statistics,'' in \emph{Proceedings Eighth IEEE
  International Conference on Computer Vision. ICCV 2001}, vol.~2.\hskip 1em
  plus 0.5em minus 0.4em\relax IEEE, 2001, pp. 416--423.

\bibitem{huang2015single}
J.-B. Huang, A.~Singh, and N.~Ahuja, ``Single image super-resolution from
  transformed self-exemplars,'' in \emph{Proceedings of the IEEE conference on
  computer vision and pattern recognition}, 2015, pp. 5197--5206.

\bibitem{matsui2017sketch}
Y.~Matsui, K.~Ito, Y.~Aramaki, A.~Fujimoto, T.~Ogawa, T.~Yamasaki, and
  K.~Aizawa, ``Sketch-based manga retrieval using manga109 dataset,''
  \emph{Multimedia Tools and Applications}, vol.~76, no.~20, pp.
  21\,811--21\,838, 2017.

\bibitem{dong2015image}
C.~Dong, C.~C. Loy, K.~He, and X.~Tang, ``Image super-resolution using deep
  convolutional networks,'' \emph{IEEE transactions on pattern analysis and
  machine intelligence}, vol.~38, no.~2, pp. 295--307, 2015.

\bibitem{kim2016accurate}
J.~Kim, J.~Kwon~Lee, and K.~Mu~Lee, ``Accurate image super-resolution using
  very deep convolutional networks,'' in \emph{Proceedings of the IEEE
  conference on computer vision and pattern recognition}, 2016, pp. 1646--1654.

\bibitem{zhang2018learning}
K.~Zhang, W.~Zuo, and L.~Zhang, ``Learning a single convolutional
  super-resolution network for multiple degradations,'' in \emph{Proceedings of
  the IEEE Conference on Computer Vision and Pattern Recognition}, 2018, pp.
  3262--3271.

\bibitem{SAN}
T.~Dai, J.~Cai, Y.~Zhang, S.-T. Xia, and L.~Zhang, ``Second-order attention
  network for single image super-resolution,'' in \emph{Proceedings of the IEEE
  conference on computer vision and pattern recognition}, 2019, pp.
  11\,065--11\,074.

\bibitem{chen2022activating}
X.~Chen, X.~Wang, J.~Zhou, and C.~Dong, ``Activating more pixels in image
  super-resolution transformer,'' \emph{CVPR}, 2023.

\bibitem{zhang2022practical}
K.~Zhang, Y.~Li, J.~Liang, J.~Cao, Y.~Zhang, H.~Tang, R.~Timofte, and
  L.~Van~Gool, ``Practical blind denoising via swin-conv-unet and data
  synthesis,'' \emph{arXiv preprint arXiv:2203.13278}, 2022.

\bibitem{zamir2022restormer}
S.~W. Zamir, A.~Arora, S.~Khan, M.~Hayat, F.~S. Khan, and M.-H. Yang,
  ``Restormer: Efficient transformer for high-resolution image restoration,''
  in \emph{Proceedings of the IEEE/CVF Conference on Computer Vision and
  Pattern Recognition}, 2022, pp. 5728--5739.

\bibitem{sheikh2005live}
H.~R. Sheikh, Z.~Wang, L.~Cormack, and A.~C. Bovik, ``Live image quality
  assessment database release 2 (2005),'' 2005.

\bibitem{foi2007pointwise}
A.~Foi, V.~Katkovnik, and K.~Egiazarian, ``Pointwise shape-adaptive dct for
  high-quality denoising and deblocking of grayscale and color images,''
  \emph{TIP}, May 2007.

\bibitem{bello2019attention}
I.~Bello, B.~Zoph, A.~Vaswani, J.~Shlens, and Q.~V. Le, ``Attention augmented
  convolutional networks,'' in \emph{Proceedings of the IEEE/CVF international
  conference on computer vision}, 2019, pp. 3286--3295.

\bibitem{zhang2015survey}
Z.~Zhang, Y.~Xu, J.~Yang, X.~Li, and D.~Zhang, ``A survey of sparse
  representation: algorithms and applications,'' \emph{IEEE access}, vol.~3,
  pp. 490--530, 2015.

\bibitem{elad2006image}
M.~Elad and M.~Aharon, ``Image denoising via sparse and redundant
  representations over learned dictionaries,'' \emph{IEEE Transactions on Image
  processing}, vol.~15, no.~12, pp. 3736--3745, 2006.

\end{thebibliography}
\end{document}